\documentclass[10pt,journal,twocolumn]{IEEEtran}
\IEEEoverridecommandlockouts
\usepackage{epsfig,cite,bm,algorithm,algorithmic,epstopdf,amsmath,subfigure, multirow,amssymb,amsfonts,amsthm,xcolor,caption,graphicx,textcomp,xcolor, subfigure, makecell, bbding, soul,lipsum}
\usepackage[inline, shortlabels]{enumitem}
\usepackage[flushleft]{threeparttable}
\soulregister{\cite}7
\soulregister{\ref}7
\setlist{itemjoin ={,\enspace},itemjoin* = { and\enspace}}

\usepackage[T1]{fontenc}
\usepackage{stfloats}
\allowdisplaybreaks[4]

\allowdisplaybreaks[4]
\usepackage{booktabs}

\newcommand{\tabincell}[2]{
\begin{tabular}{@{}#1@{}}#2\end{tabular}}

\begin{document}

\title{Rate-Splitting Multiple Access for  Secure Near-Field Integrated Sensing and Communication}
%\title{Scure Hybrid Beamfocusing for RSMA-enhanced Near-Field Integrated Sensing and Communication}
	
\author{Jiasi Zhou, Chintha Tellambura,~\IEEEmembership{Fellow,~IEEE}, and Geoffrey Ye Li,~\IEEEmembership{Fellow,~IEEE}
\thanks{Jiasi Zhou is with the School of Medical Information and Engineering, Xuzhou Medical University, Xuzhou, 221004, China, (email: jiasi\_zhou@xzhmu.edu.cn). (\emph{Corresponding author: Jiasi Zhou}).}
\thanks{ Chintha Tellambura is with the Department of Electrical and Computer Engineering, University of Alberta, Edmonton, AB, T6G 2R3, Canada (email: ct4@ualberta.ca).}
\thanks{Geoffrey Ye Li is with the School of Electrical and Electronic Engineering, Imperial College London, London SW7 2AZ, UK (e-mail: geoffrey.li@imperial.ac.uk).}
\thanks{This work was supported by the Talented Scientific Research Foundation of Xuzhou Medical University (D2022027) and Basic Science of Higher Education Institutions in Jiangsu Province (25KJB510034).}}
\maketitle
\begin{abstract}
Near-field integrated sensing and communication (ISAC) leverages distance-dependent channel variations for joint distance–angle estimation. However, full-digital architectures have prohibitive hardware costs, making hybrid analog–digital (HAD) designs the primary alternative. Nevertheless, such architectures compromise beamfocusing precision and lead to energy leakage, which exacerbates inter-user interference and heightens eavesdropping risks. To address these challenges, this paper proposes a rate-splitting multiple access (RSMA)-enhanced secure transmit scheme for near-field ISAC. For the first time, it exploits the common stream in  RSMA to concurrently (i) flexibly manage interference, (ii) act as artificial noise to suppress eavesdropping, and (iii) serve as sensing sequences. The objective is to maximize the minimum secrecy rate while satisfying the angle and distance Cram\'{e}r–Rao bound (CRB) constraints. This results in a hard,  non-convex optimization problem, and we employ block coordinate descent to decompose it into three sub-problems with lower computational complexity. In the first stage of optimizing fully digital beamfocusers, we develop an iterative solution using weighted minimum mean-squared error (WMMSE), quadratic transform, and Taylor expansion methods — thus avoiding conventional semidefinite relaxation (SDR). In the second and third stages,  the analog and digital beamfocusers are optimized in closed form. Simulation results show that the proposed scheme (1) achieves near fully digital beamfocusing performance but with a 16-fold reduction in RF chains, (2) provides superior secrecy performance compared to conventional beamfocusing-only and far-field security schemes, and (3) enables high-accuracy sensing with negligible loss in secrecy performance.
\end{abstract} 


\section{Introduction} 
Integrated sensing and communication (ISAC) unifies radar sensing and data transmission within a single framework, thereby enhancing spectral efficiency and reducing hardware complexity~\cite{9737357}. To meet increasingly stringent and diverse service demands, future ISAC networks are expected to operate in high-frequency bands, such as millimeter-wave (mmWave) and Terahertz (THz) spectra~\cite{10422722}. However, these high-frequency signals suffer from severe path loss, which degrades both communication throughput and sensing reliability. To compensate for this propagation loss, extremely large-scale antenna arrays (ELAAs) are exploited.

Operating with ELAAs at such high frequencies pushes ISAC systems into the near-field propagation regime, where the traditional plane-wave model becomes invalid, giving way to a distance-dependent spherical-wavefront model~\cite{10220205}. Unlike plane waves, spherical wavefronts vary across both angle and distance dimensions. This property enables joint distance–angle estimation for high-precision target sensing~\cite{he2024unlocking}. Moreover, it facilitates spotlight-like beamfocusing, where radiated energy is confined to specific spatial regions, effectively suppressing multi-user interference in communication~\cite{an2023toward}. These capabilities open new opportunities for advanced near-field ISAC design.

Fully digital architectures require one radio-frequency (RF) chain per antenna, resulting in high hardware costs and power consumption that scale directly with array size. In near-field scenarios with ELAAs, this scaling becomes prohibitive, rendering such architectures impractical~\cite{11015468}.
As a practical alternative, hybrid analog–digital (HAD) architectures integrate fewer digital baseband chains with analog phase shifters to cooperatively control beamfocusing~\cite{7010533}. However, it introduces two critical technical challenges, as outlined below.

\begin{itemize}
\item \textbf{Interference Control:} 
Unlike conventional communication networks, ISAC systems inherently introduce cross-functional interference because communication and sensing share the same spectrum~\cite{10486996}. For instance, communication beams with highly concentrated energy narrow the sensing coverage region and degrade target detection performance~\cite{11129050}. Mitigating this loss typically requires adding dedicated sensing beams, but these beams in turn create additional interference for communication links. As a result, precise cross-functional interference control becomes critical for achieving high communication throughput and accurate sensing. However, the low-dimensional analog beams in HAD architectures cannot provide fine-grained beamfocusing, resulting in unavoidable energy leakage and elevated interference levels. Recent work~\cite{zhou2025sub} has demonstrated that near-field HAD beamfocusing alone cannot adequately suppress this interference, underscoring the need for more advanced interference management strategies.

\item \textbf{Privacy Security:}
HAD architectures inherently limit beamfocusing precision, causing unavoidable energy leakage into unintended spatial regions~\cite{zhou2025sub}. While this dispersed energy can aid target sensing, it also elevates eavesdropping risks due to the broadcast nature of wireless channels~\cite{9927490}. When eavesdroppers are located near legitimate users, hybrid beamfocusing alone cannot guarantee confidentiality, necessitating additional measures, such as artificial noise injection~\cite{10480457}. These safeguards, however, introduce trade-offs, including increased power consumption and reduced effective capacity for legitimate users.
\end{itemize}

Rate-splitting multiple access (RSMA) offers an integrative solution to address these challenges. The key idea of RSMA is that at the transmitter side, the base station (BS) splits each message into a common part and a private part~\cite{mao2018rate,9831440}. The BS jointly encodes all common parts into a common stream, whereas each private part is encoded into a user-specific stream. At the receiver side, each user decodes the common stream, removes it from the received signal via successive interference cancellation (SIC), and then decodes its private stream. The common stream can be deliberately designed to fulfill three functions, summarized as follows:
\begin{itemize}
\item \textbf{Flexible Interference Manager:} 
The common stream enables flexible interference mitigation by adjusting the message-splitting ratio, allowing each user to decode and remove part of the interference while tolerating the small residual. This tunable structure significantly enhances the flexibility and robustness of multi-user interference management, with SDMA and NOMA emerging as special cases~\cite{9991090,10038476}. Furthermore, in ISAC systems, the common stream provides an additional degree of freedom to coordinate interference between communication and sensing functions, thereby enabling effective cross-functional interference management~\cite{9531484}.
\item \textbf{Dedicated Sensing Sequence:} The common stream can also be optimized to serve as a dedicated sensing sequence for accurate target detection, which eliminates the need for separate radar sequences~\cite{9531484,11129050}.
\item \textbf{Artificial Noise source:} In eavesdropping scenarios, the common stream can also function as artificial noise, degrading the eavesdropper’s reception while still conveying useful information to legitimate users~\cite{10945425,zhang2025fluid}. This dual role eliminates the need for dedicated artificial noise and can substantially enhance secrecy performance, including improvements in both the max–min secrecy rate and the secrecy sum rate.
\end{itemize}
Existing near-field ISAC studies have explored the common stream to optimize only the first two functions, achieving considerable gains in communication throughput and sensing precision~\cite{11129050,zhou2025crb}. However, to our knowledge, no previous work has exploited secure RSMA-enabled near-field ISACs or leveraged the common stream to optimize all three functions,  which motivates our work.

\subsection{Related Works}
\subsubsection{Securing near-field communications/ISAC} Extensive research has been conducted on securing near-field communications and ISAC~\cite{10436390,10971913,liu2025physical,zhang2024near,10902048,zhao2025near,ren2024secure,11018844}. Near-field beamfocusing has been confirmed to enhance secrecy rate~\cite{10436390} and enlarges the secure region~\cite{10971913} compared with far-field beamforming, particularly when eavesdroppers and intended receivers are aligned in the same direction. In contrast, a pronounced contraction of the secure region is reported in~\cite{liu2025physical} when eavesdroppers operate in the near-field while legitimate users remain in the far-field. To address security challenges in wideband systems, true-time delayer (TTD)-based analog beamfocusing is explored to mitigate the beam-split effect~\cite{zhang2024near}. Moreover, beamfocusing can be integrated with far-to-near SIC schemes to prevent privacy leakage of far-end users~\cite{10902048}. Besides, secure near-field ISAC has also been investigated in~\cite{zhao2025near} and~\cite{ren2024secure} while the sensing capability of ISAC  is exploited in~\cite{11018844} to counteract mobile eavesdroppers. However, these publications inject dedicated sensing beams to act as artificial noise~\cite{zhao2025near,ren2024secure,11018844}, which consumes valuable power that could otherwise support data communication.

\subsubsection{RSMA-enhanced securing far-field ISAC} To effectively manage multi-user interference, an RSMA-based secure transmit scheme is designed for far-field ISAC in~\cite{li2025secure}. This scheme can be further integrated with active reconfigurable intelligent surfaces (RISs) to address the spatial uncertainty of eavesdroppers\cite{salem2024robust}. When the channel state information (CSI) of eavesdroppers is unavailable, artificial noise can be introduced as an additional component to disrupt potential eavesdropping~\cite{dan2025beamforming}. To improve energy efficiency, the dual functionality of the common stream is exploited in~\cite{10847905,zhang2025fluid,10812007}, thereby eliminating the necessity for artificial noise. Under imperfect CSI conditions, fluid antenna~\cite{zhang2025fluid} or passive RIS\cite{10812007} can be incorporated to enhance network robustness. However, these solutions are built upon the plane wave assumption and fully digital antenna architectures, which may limit their applicability in near-field scenarios.

\subsubsection{RSMA-enabled near-field communications/ISAC} To better align with practical propagation environments, near-field RSMA has been preliminarily investigated in~\cite{11071287,zhou2025sub,10798456,10414053,10906379,11129050,zhou2025crb}. In particular, RSMA-based transmit schemes under imperfect CSI and SIC conditions in~\cite{zhou2025sub} and~\cite{11071287} demonstrate superior interference management capability. This superiority suggests that hybrid beamfocusing induces energy leakage and, therefore, cannot eliminate multi-user interference. The leaked energy can be harnessed to support additional communication users~\cite{10798456} or sense near-field targets~\cite{10906379}, yet the associated non-negligible eavesdropping risks remain unaddressed. In parallel,  a more realistic mixed far-field and near-field scenario is investigated in~\cite{10414053}, which employs a fully-connected hybrid antenna architecture to reduce hardware cost. Building on similar transmit frameworks, the common stream is leveraged to coordinate interference between communication and sensing functions~\cite{11129050}.  However, the adopted sensing metric fails to directly capture target location information. To fully exploit the embedded information in the spherical wave channel modeling, the Cram\'{e}r–Rao bound (CRB) is employed as a more effective sensing metric in~\cite{zhou2025crb}.

\subsection{Main Contributions}
Based on the above discussion, novel solutions are needed to address the interference and security challenges inherent in HAD-based NF-ISAC systems. This paper is the first to exploit the triple functionality of the RSMA common stream to tackle these issues. As summarized in Table \ref{Table I}, existing works cover only isolated components—near-field modeling, ISAC signaling, physical-layer security, or RSMA—whereas our framework uniquely integrates all four within a single design. Furthermore, prior RSMA-based studies typically utilize only one or two RSMA functionalities (interference management, sensing-sequence embedding, or artificial-noise generation). In contrast, our approach leverages all three, enabling precise cross-functional interference control, flexible sensing-sequence design, and inherent secrecy enhancement without the need for dedicated artificial noise. As a result, our scheme is the only one that satisfies all the criteria in the comparison, offering a comprehensive and novel solution for secure near-field ISAC.

However, the formulated max-min secrecy rate optimization problem is non-convex and intractable with common convex algorithms. To tackle this challenge, prior studies primarily depend on semi-definite relaxation (SDR) techniques~\cite{zhou2025crb}, where rank-one auxiliary matrices are introduced (See~\cite{ren2024secure} for details). Under this framework, the angle and distance CRB constraints can be reformulated into convex sets using the Schur complement, thereby enabling tractable optimization. However, SDR-based algorithms exhibit two critical limitations. First, the computational complexity becomes prohibitive in ELAAs or multi-user scenarios. Second, the rank-one constraint cannot be guaranteed, necessitating randomization or approximation procedures that ultimately degrade network performance. To address these issues, we propose a penalty-based block coordinate descent (BCD) algorithm that directly optimizes the beamfocusing vectors, effectively avoiding the drawbacks of conventional SDR approaches. The detailed contributions are summarized as follows:
\begin{table*}[h]
	\caption{Our contributions in comparison to the related works}
	\begin{center}\label{Table I}
	\begin{tabular}{|c|c|c|c|c|c|c|c|} 
\hline
\multicolumn{2}{|c|}{} &\cite{10436390,10971913,liu2025physical,zhang2024near,10902048,zhao2025near,ren2024secure,11018844} &\cite{li2025secure,10847905,zhang2025fluid,10812007}&\cite{salem2024robust,dan2025beamforming}&\cite{zhou2025sub,11071287,10798456,10906379}&\cite{zhou2025crb,11129050}&\makecell*[c]{\bf{Our work}}\\  
\hline
\multicolumn{2}{|c|}{\bf{Near-field}}&\checkmark &&&\checkmark&\checkmark&\checkmark\\  
\hline
\multicolumn{2}{|c|}{\bf{ISAC}}&\checkmark &\checkmark&\checkmark&&\checkmark&\checkmark\\
\hline
\multicolumn{2}{|c|}{\bf{Physical layer security}}&\checkmark &\checkmark&\checkmark&&&\checkmark\\
\hline
\multirow{3}*{\tabincell{c}{\bf{Threefold}\\{\bf{functionality}}\\{\bf{of RSMA}}}}&Interference management&&\checkmark&\checkmark&\checkmark&\checkmark&\checkmark\\ 
\cline{2-8}  
&Sensing sequences&&&\checkmark&&\checkmark&\checkmark\\
\cline{2-8}
&Artificial noise&&\checkmark&&&&\checkmark\\
\hline
\end{tabular}
	\end{center}
\end{table*}

\begin{itemize}
\item \textbf{Novel Secure Scheme for Near-Field ISAC:} The transmit framework exploits the RSMA common stream to achieve three concurrent purposes: flexibly managing interference, acting as artificial noise against eavesdropping, and providing sensing sequences. The angle and distance CRBs are derived, and a minimum secrecy rate maximization problem is formulated under these CRB constraints. This formulation entails the joint optimization of the analog beamfocuser, digital beamfocuser, and common secrecy rate allocation.

\item \textbf{Efficient Algorithm Design:} To solve the resulting non-convex problem, we introduce fully digital beamfocusers as auxiliary variables and then propose a penalty-based BCD algorithm. Within this framework, the optimization variables are partitioned into three blocks and updated alternately, guaranteeing convergence to a stationary solution.
\begin{enumerate}
\item \emph{Fully digital beamfocusing}: The legitimate rate, eavesdropping rate, and sensing CRBs are reformulated via weighted minimum mean-squared error (WMMSE), quadratic transform, and first-order Taylor expansion, respectively. An iterative algorithm is then developed to solve the subproblem without resorting to conventional SDR.  
\item \emph{Analog beamfocusing:} This subproblem proceeds on an element-by-element basis, where each entry admits a closed-form update.
\item \emph{Digital beamfocusing:} The optimal digital beamfocuser is obtained in closed form based on the first-order optimality conditions.
\end{enumerate}

\item \textbf{Performance Evaluation:} Simulation results verify that our transmit scheme: (1) achieves near fully digital beamfocusing performance while requiring substantially fewer RF chains, (2) delivers superior secrecy performance compared with conventional beamfocusing-only and far-field security approaches, and (3) attains high-accuracy sensing with negligible degradation in communication quality.
\end{itemize}

\emph{Organization:}  The rest of this paper is organized as follows.  Section \ref{Section II} introduces the system model and formulates the secrecy rate maximization problem. Section \ref{Section III} presents an efficient iterative optimization algorithm and analyzes its key properties. Section \ref{Section IV} provides simulation results to evaluate our algorithm. Finally, section \ref{Section V} concludes this paper.

\emph{Notations:} Boldface upper-case letters, boldface lower-case letters, and calligraphy letters denote matrices, vectors, and sets. The complex matrix space of $N\times K$ dimensions is denoted by $\mathbb{C}^{N\times K}$. Superscripts ${(\bullet)}^T$, ${(\bullet)}^H$, and ${(\bullet)}^\dagger$ represent the transpose, Hermitian transpose, and pseudo-inverse, respectively. $\text{Re}\left( \bullet\right)$, $\text{Tr}\left( \bullet\right)$, and $\mathbb{E}\left[\bullet\right]$ 
 denote the real part, trace, and statistical expectation. The Frobenius norm of matrix $\mathbf{X}$ is denoted by $||\mathbf{X}||_F$. $\odot$ and $\otimes$ denote the Hadamard and Kronecker products. Variable $x\sim\mathcal{CN}(\mu, \sigma^2)$ is a  circularly symmetric complex Gaussian (CSCG) with mean $\mu$ and variance $\sigma^2$. 

\section{System Model and Problem Formulation}\label{Section II}
\begin{figure}[tbp]
\centering
\includegraphics[scale=0.57]{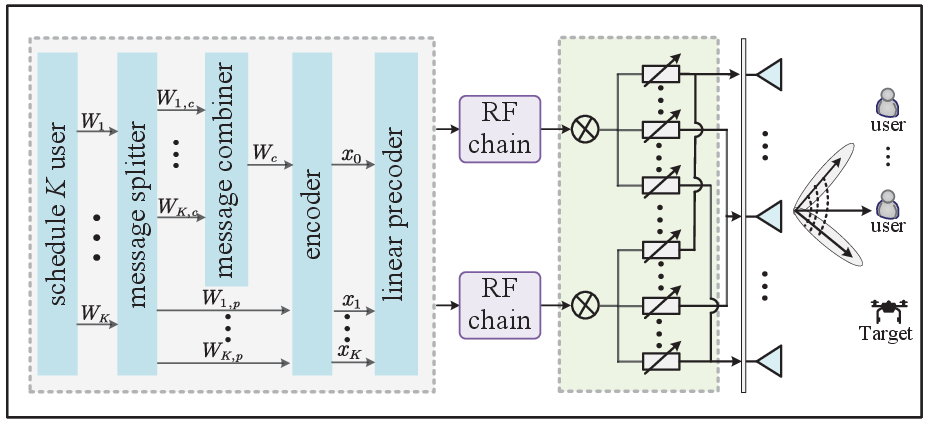}
\caption{RSMA-enhanced secure near-field ISAC networks.}
\label{system}
\end{figure}
As illustrated in Fig.~\ref{system}, we consider an RSMA-enhanced secure near-field ISAC, which comprises a dual-functional BS, $K$ single-antenna communication users, and one target acting as a potential eavesdropper. Both the communication users and the target are assumed to lie within the near-field region. To reduce power consumption and hardware complexity, a fully-connected HAD beamfocusing architecture is adopted, wherein each RF chain is connected to all transmit antennas. The BS is equipped with a uniform linear array (ULA) of $N$ transmit antennas, $M$ receive antennas, and $L$ RF chains, with an inter-antenna spacing denoted by $d$. Let $\mathcal{K}\triangleq\{1,\dots, K\}$, $\mathcal{N}\triangleq\{1,\dots, N\}$, $\mathcal{M}\triangleq\{1,\dots, M\}$, and $\mathcal{L}\triangleq\{1,\dots, L\}$ represent the sets of communication users, transmit antennas, receive antennas, and RF chains, respectively.

\subsection{Near-field communication and sensing channel models}
Without loss of generality, we assume that the transmit and receive antennas are placed along the positive and negative $y$-axis, respectively. Accordingly, the coordinate of the $n$-transmit antenna is $\mathbf{t}_n= (0,nd)$ for $\forall n\in\mathcal{N}$. Let $r_k$ and $\theta_k$ denote the distance and angle of the $k$-th communication user with respect to the origin, whose coordinate is $\mathbf{u}_k = (r_k\cos\theta_k, r_k\sin\theta_k)$. The propagation distance between transmit antenna $n$ and user $k$ is given by
\begin{align}
d_{n,k} &= \left\|\mathbf{t}_n-\mathbf{u}_k\right\| = \sqrt{r^2_k + (nd)^2 - 2ndr_k \sin\theta_k}\notag\\
&\overset{(a)}{\approx} r_k-nd\sin\theta_k+\frac{(nd)^2\cos^2\theta_k}{2r_k},
\end{align}
where $\overset{(a)}{\approx}$ comes from the second-order Taylor expansion. This approximation characterizes amplitude- and phase-dependent variations across channel links between each transmit element and the $k$-th communication user. Following~\cite{10220205}, the $n$-th element of near-field array response vector $\mathbf{a}\left(r_k,\theta_k\right)\in\mathbb{C}^{N\times 1}$ can be expressed as 
\begin{align}
\left[\mathbf{a}\left(r_k,\theta_k\right)\right]_n=e^{-j\frac{2\pi}{\lambda}\left(-nd\sin\theta_k+\frac{(nd)^2\cos^2\theta_k}{2r_k}\right)}.
\end{align}
Meanwhile, the $n$-th element of the complex channel gain vector ${\boldsymbol{\beta}}_{k}\in\mathbb{C}^{N\times 1}$ can be given by 
\begin{align}
\left[{\boldsymbol{\beta}}_{k}\right]_n=\frac{c}{4\pi f d_{n,k}}e^{-j\frac{2\pi}{\lambda}r_k}\overset{(a)}{\approx}\frac{c}{4\pi f r_k}e^{-j\frac{2\pi}{\lambda}r_k}.
\end{align}
where $c$, $f$, and $\lambda$ are the speed of light, carrier frequency, and wavelength, respectively. $\overset{(a)}{\approx}$ follows from the Fresnel approximation, which reveals that the path loss across all transmit antennas is approximately identical.

By combining the array response and complex channel gain vectors, the channel $\mathbf{h}_k\in\mathbb{C}^{N\times 1}$ between the transmit antenna array and the $k$-th communication user can be modeled as
\begin{align}
\mathbf{h}_k={\boldsymbol{\beta}}_{k}\odot\mathbf{a}\left(r_k,\theta_k\right)\approx \beta_k\mathbf{a}\left(r_k,\theta_k\right),
\end{align}
where $\beta_k=\frac{c}{4\pi f r_k}e^{-j\frac{2\pi}{\lambda}r_k}$. Similarly, the eavesdropping channel vector can be given by $\mathbf{g}_\text{e}= \beta_\text{e}\mathbf{a}\left(r_\text{e},\theta_\text{e}\right)\in\mathbb{C}^{N\times 1}$, where $\left(r_\text{e},\theta_\text{e}\right)$ are the polar coordinates of the near-field target and $\beta_\text{e}$ is the corresponding complex channel gain.

To estimate the target's parameters, the BS transmits probing signals toward the target and collects the corresponding echo signals, thereby introducing a round-trip propagation channel~\cite{11129050}. Analogous to the communication channel model, the sensing channel matrix $\mathbf{G}\in\mathbb{C}^{M\times N}$ is 
\begin{equation}
\mathbf{G}=\tilde\beta_\text{e}\mathbf{b}\left(r_\text{e},\theta_\text{e}\right)\mathbf{a}^T\left(r_\text{e},\theta_\text{e}\right),
\end{equation}
where $\tilde \beta_\text{e}$ denotes the complex channel gain and $\mathbf{b}\left(r_\text{e},\theta_\text{e}\right)\in\mathbb{C}^{M\times 1}$ represents the receive array response vector associated with the target.

\subsection{RSMA-enhanced signal transmission model}
\subsubsection{Communication performance} Let $T$ denote discrete-time indices within one coherent processing interval (CPI). At the $t$-th time index for $\forall t\in\mathcal{T}=\{1,\dots, T\}$, the message $W_k(t)$ for the $k$-th communication user is divided into a private part $W_{k,p}(t)$ and a common part $W_{k,c}(t)$. The private part $W_{k,p}(t)$ is encoded into a private stream $x_k(t)$ with unit power, i.e., $\mathbb{E}|x_{k}(t)|^2=1$, for $\forall k\in\mathcal{K}$. All common parts are combined into a common message, i.e., $W_0(t)=\left\{W_{1,c}(t),\dots,W_{K,c}(t)\right\}$, which is encoded into a common stream $x_0(t)$ and superimposed on the top of the private streams. Consequently, the aggregated stream vector at time index $t$ is expressed as $\mathbf{x}\left(t\right)=\left[x_0(t),x_1(t),\dots,x_K(t)\right]^T\in\mathbb{C}^{(K+1)\times 1}$. This stream vector is linearly precoded by a fully-connected hybrid beamfocuser $\mathbf{FW}\in\mathbb{C}^{N\times (K+1)}$, so the transmitted signal at the $t$-th time index is
\begin{align}
\tilde{\mathbf{x}}(t)=\mathbf{FWx}(t),
\end{align}
where $\mathbf{F}\in\mathbb{C}^{N\times M}$ and $\mathbf{W}=\left[\mathbf{w}_0,\dots, \mathbf{w}_K\right]\in\mathbb{C}^{M\times (K+1)}$ represent analog and digital beamfocusing matrices, respectively. In particular, $\mathbf{w}_k\in\mathbb{C}^{L\times 1}$ denotes digital beamfocusing vector associated with stream $x_k(t)$.  Let $\mathbf{R}=\mathbb{E}\left[\tilde{\mathbf{x}}(t)\tilde{\mathbf{x}}^H(t)\right]$ denote the covariance matrix of the sensing signal, so we have $ \mathbf{R}=\mathbf{FW}\mathbf{W}^H\mathbf{F}^H$. Under the fully-connected configuration, each element of analog beamfocuser $\mathbf{F}$ must satisfy the unit-modulus constraint, i.e.,
\begin{align}
\mathbf{F}_{n,m}=\left\{e^{j\theta}|\theta\in\left(0,2\pi\right]\right\},~\forall n\in\mathcal{N},~\forall m\in\mathcal{M}.
\end{align}

With this process, the received signal by the $k$-th communication user and the target at time index $t$ are respectively expressed as
\begin{subequations}
\begin{align}
y_{k}(t)=&\mathbf{h}^H_k\mathbf{F}\mathbf{w}_{0}x_{0}(t)+ \sum\nolimits^K_{i=1}\mathbf{h}^H_k\mathbf{F}\mathbf{w}_{i}x_{i}(t)+n_{k},\\
y_{\text{e}}(t)=&\mathbf{g}^H_{\text{e}}\mathbf{F}\mathbf{w}_{0}x_{0}(t)+ \sum\nolimits^K_{i=1}\mathbf{g}^H_{\text{e}}\mathbf{F}\mathbf{w}_{i}x_{i}(t)+n_{\text{e}},
\end{align}
\end{subequations}
where $n_{k}\sim \mathcal{CN}\left(0,\sigma^2_{k}\right)$ and $n_{\text{e}}\sim \mathcal{CN}\left(0,\sigma^2_{\text{e}}\right)$ represent additional white Gaussian noise (AWGN). To facilitate the interpretation of the achievable rates for both legitimate users and the eavesdropper, we next provide the average received power at the $k$-th communication user and the target, which are respectively given by~\cite{7555358}
\begin{subequations}\label{power}
	\begin{align}
&T_{k,c}=\overbrace{{\left|\mathbf{h}^H_{k}\mathbf{Fw}_0\right|}^2}^{S_{k,c}}+\underbrace{\overbrace{{\left|\mathbf{h}^H_{k}\mathbf{Fw}_{k}\right|}^2}^{S_{{k,p}}}+\overbrace{\sum_{i=1,i\neq k}^{K}\left|\mathbf{h}^H_{k}\mathbf{Fw}_i\right|^2+\sigma^2_k}^{I_{{k,p}}}}_{I_{k,c}=T_{k,p}},
	\label{equ:power_legitimate}\\
&T_{\text{e},c}=\overbrace{{\left|\mathbf{g}^H_\text{e}\mathbf{Fw}_0\right|}^2}^{S_{\text{e},c}}+\underbrace{\overbrace{{\left|\mathbf{g}^H_\text{e}\mathbf{Fw}_{k}\right|}^2}^{S_{{\text{e},k}}}+\overbrace{\sum_{i=1,i\neq k}^{K}\left|\mathbf{g}^H_\text{e}\mathbf{Fw}_i\right|^2+\sigma^2_\text{e}}^{I_{{\text{e},k}}}}_{I_{\text{e},c}=T_{\text{e},k}}.
	\label{equ:power_eavesdropper}
	\end{align}
\end{subequations}

For the proposed RSMA-based secure transmit scheme, each communication user first decodes the common stream by treating all private streams as interference and then extracts its intended common message. Afterward, the common stream is removed via SIC, after which the private stream is decoded. Accordingly, the signal-to-interference-plus-noise ratios (SINRs) of the common and private streams at the $k$-th communication user are expressed as
\begin{equation}
\gamma_{k,c}=\frac{S_{k,c}}{I_{k,c}}~\text{and}~\gamma_{k,p}=\frac{S_{k,p}}{I_{k,p}}.
\label{Legiti_SINR}
\end{equation}
Under Gaussian signaling, the achievable rates (bps/Hz) for decoding the common and private streams are respectively given by
\begin{equation}
R_{k,c}=\log\left(1 + \gamma_{{k,c}}\right)~\text{and}~R_{k,p}=\log\left(1 + \gamma_{{k,p}}\right).\label{Legit}
\end{equation}
To guarantee that the common stream is decodable by all communication users, its actual achievable rate $R_c$ must satisfy $R_c=\min_{\forall k\in\mathcal{K}}R_{k,c}$. 

In parallel, the sensing target attempts to eavesdrop on both the private and common streams. When the common stream can serve as artificial noise, the SINRs for decoding the common and private streams at the target are given by
\begin{equation}
\gamma_{\text{e},c}=\frac{S_{\text{e},c}}{I_{\text{e},c}} ~\text{and}~\gamma_{\text{e},k}=\frac{S_{\text{e},k}}{I_{\text{e},k} + S_{\text{e},c}}.
\label{Eaves_SINR}
\end{equation}
The corresponding eavesdropping rates are respectively expressed as
\begin{equation}
R_{\text{e},c}=\log\left(1 + \gamma_{{\text{e},c}}\right)~\text{and}~R_{\text{e},k}=\log\left(1+\gamma_{\text{e},k}\right).\label{eaves}
\end{equation}
However, for the common stream to effectively serve as artificial noise, eavesdropping rate $R_{\text{e},c}$ should be lower than the actual common rate, i.e., $R_{\text{e},c}\leq R_c$. Based on (\ref{Legit}) and (\ref{eaves}), the secrecy rates of the common and private streams are respectively given by
\begin{equation}
R^s_{c}=\left[R_c-R_{\text{e},c}\right]^+~\text{and}~R^s_{k,p} = {[R_{k,p}-R_{\text{e},k}]}^+,
\end{equation}
where ${[x]}^+=\max{\left(x,0\right)}$. Moreover, since all communication users share the secrecy rate of the common stream, it follows that $\sum^{K}_{k=1}R^s_{k,c}\leq R^s_{c}$, where $R^s_{k,c}$ denotes the portion of the common secrecy rate allocated to the $k$-th communication user. Consequently, the total secrecy rate achievable by the $k$-th communication user is expressed as
\begin{equation}
    R^s_{k} = R^s_{k,c} + R^s_{k,p}. 
\label{ssr}
\end{equation} 

\subsubsection{Sensing CRB} This paper focuses on target localization by estimating distance $r_{\text{e}}$ and angle $\theta_{\text{e}}$. To this end, the sensing performance is typically evaluated using the mean-squared error (MSE) between the estimated and real coordinates. However, deriving a closed-form expression for the MSE is intractable. As a practical alternative, we adopt the CRB as the performance metric for target sensing, which provides a closed-form expression to characterize the lower bound of the achievable MSE. Let $\mathbf{X}=\left[\tilde{\mathbf{x}}(1),\dots,\tilde{\mathbf{x}}(T)\right]\in\mathbb{C}^{N\times T}$ and $\mathbf{Y}\in\mathbb{C}^{N\times T}$ denote the accumulated transmitted signal and received echo signal over $T$ time slots. The received echo signal $\mathbf{Y}$ at the BS can be expressed as
\begin{equation}
\mathbf{Y}=\tilde\beta_\text{e}\mathbf{b}\left(r_\text{e},\theta_\text{e}\right)\mathbf{a}^T\left(r_\text{e},\theta_\text{e}\right)\mathbf{X}+\mathbf{N}_0,
\label{Sensing_matrix}
\end{equation}
where $\mathbf{N}_0\in\mathbb{C}^{N\times T}$ denotes the AWGN matrix, with each entry being a zero-mean CSCG random variable with variance $\sigma^2_{\text{e}}$. Then, $\mathbf{Y}$ can be vectorized as  
\begin{equation}
\mathbf{y}=\text{vec}\left(\mathbf{Y}\right)=\mathbf{u} +\tilde{\mathbf{n}},
\label{Vectorized_Sensing}
\end{equation}
where $\mathbf{u}=\text{vec}\left(\tilde\beta_\text{e}\mathbf{b}\left(r_\text{e},\theta_\text{e}\right)\mathbf{a}^T\left(r_\text{e},\theta_\text{e}\right)\mathbf{X}\right)\in\mathbb{C}^{NT\times 1}$ and $\tilde{\mathbf{n}}=\text{vec}\left(\mathbf{N}_0\right)\sim \mathcal{CN}\left(\mathbf{0},\mathbf{v}\right)$ with $\mathbf{v}=\sigma^2_{\text{e}}\mathbf{I}_{NT}$. According to the assumptions of zero-mean complex Gaussian noise, the complex Gaussian vector follows $\mathbf{y}\sim \mathcal{CN}\left(\mathbf{u},\mathbf{v}\right)$.

We define unknown parameter vector as $\boldsymbol{\xi}=\left[r_{\text{e}},\theta_{\text{e}},\tilde\beta^{\text{R}}_{\text{e}},\tilde\beta^{\text{I}}_{\text{e}}\right]$, where $\tilde\beta^{\text{R}}_{\text{e}}$ and $\tilde\beta^{\text{I}}_{\text{e}}$ denote the real and imaginary parts of $\tilde\beta_{\text{e}}$, respectively. The Fisher information matrix (FIM) for estimating parameters $\bm{\xi}$ can be partitioned as
\begin{equation}
\mathbf{J}_{\bm{\xi}}= \frac{2T}{\sigma^2_{\text{e}}}\begin{bmatrix} \mathbf{J}_{11} & \mathbf{J}_{12} \\ \mathbf{J}^T_{12} & \mathbf{J}_{22} \end{bmatrix}\in\mathbb{R}^{4\times 4},
\label{FIM_a}
\end{equation}
Matrices $\mathbf{J}_{11}$, $\mathbf{J}_{12}$, and $\mathbf{J}_{22}$ in (\ref{FIM_a}) are derived in Appendix A and can be expressed as. 
\begin{subequations}\label{FIM_3}
	\begin{align}
 &\mathbf{J}_{11}= |\tilde\beta_{\text{e}}|^2\text{Re}\left(\begin{bmatrix} \text{Tr}\left(\dot{\mathbf{G}}_{\theta_{\text{e}}}\mathbf{R}\dot{\mathbf{G}}^H_{\theta_{\text{e}}}\right) & \text{Tr}\left(\dot{\mathbf{G}}_{r_{\text{e}}}\mathbf{R}\dot{\mathbf{G}}^H_{\theta_{\text{e}}}\right) \\ \text{Tr}\left(\dot{\mathbf{G}}_{\theta_{\text{e}}}\mathbf{R}\dot{\mathbf{G}}^H_{r_{\text{e}}}\right) & \text{Tr}\left(\dot{\mathbf{G}}_{r_{\text{e}}}\mathbf{R}\dot{\mathbf{G}}^H_{r_{\text{e}}}\right) \end{bmatrix}\right),\\
&\mathbf{J}_{12}=\text{Re}\left(\begin{bmatrix} \tilde{\beta}^*_{\text{e}}\text{Tr}(\tilde{\mathbf{ G}}\mathbf{R}\dot{\mathbf{G}}^H_{\theta_{\text{e}}}) \\ \tilde{\beta}^*_{\text{e}}\text{Tr}(\tilde{\mathbf{ G}}\mathbf{R}\dot{\mathbf{G}}^H_{r_{\text{e}}}) \end{bmatrix}\begin{bmatrix} 1,j \end{bmatrix}\right),\\
 &\mathbf{J}_{22}=\mathbf{I}_2\text{Tr}\left(\tilde{\mathbf{ G}}\mathbf{R}\tilde{\mathbf{ G}}^H\right).
	\end{align}
\end{subequations}
where $\tilde{\mathbf{G}}=\mathbf{b}\left(r_\text{e},\theta_\text{e}\right)\mathbf{a}^T\left(r_\text{e},\theta_\text{e}\right)$, $\dot{\mathbf{G}}_{r_\text{e}} = \frac{\partial \tilde{\mathbf{G}}}{\partial r_{\text{e}}}$, and $\dot{\mathbf{G}}_{\theta_\text{e}} = \frac{\partial \tilde{\mathbf{G}}}{\partial \theta_{\text{e}}}$. Based on (\ref{FIM_a}), the CRB matrix for jointly estimating distance and angle can be given by 
\begin{equation}
\text{CRB}\left(r_{\text{e}},\theta_{\text{e}}\right) = \frac{\sigma^2_0}{2T}\left(\mathbf{J}_{11}-\mathbf{J}_{12}\mathbf{J}^{-1}_{22}\mathbf{J}^T_{12}\right)^{-1}\in\mathbb{R}^{2\times 2}.\label{CRB}
\end{equation} 
More specifically, the MSEs of distance and angle are bounded as $\epsilon^2_{\theta_{\text{e}}}\geq [\text{CRB}\left(r_{\text{e}},\theta_{\text{e}}\right)]_{1,1}$ and $\epsilon^2_{r_{\text{e}}}\geq [\text{CRB}\left(r_{\text{e}},\theta_{\text{e}}\right)]_{2,2}$, respectively.

\subsection{Problem formulation}
This paper aims to maximize the minimum secrecy rate by jointly designing the fully-connected hybrid beamfocuser and the common secrecy rate while ensuring target localization accuracy and transmit power requirements. The resultant optimization problem can be formulated as
\begin{subequations}\label{linear_p}
	\begin{align}
&\max_{\mathbf{F},\mathbf{W},R^s_{k,c} } \min_{\forall k} R^s_k,\label{ob_a}\\
	\text{s.t.}~
&[\text{CRB}\left(r_{\text{e}},\theta_{\text{e}}\right)]_{1,1}\leq\Gamma_{\theta_{\text{e}}},\label{ob_b}\\
&[\text{CRB}\left(r_{\text{e}},\theta_{\text{e}}\right)]_{2,2}\leq\Gamma_{r_{\text{e}}},\label{ob_c}\\
&||\mathbf{FW}||^2_F\leq P_{\text{th}},\label{ob_d}\\
&\min_{\forall k\in\mathcal{K}}R_{k,c}\geq R_{\text{e},c},\label{ob_e}\\
&\sum_{k=1}^{K}R^s_{k,c} \leq R^s_{c},\label{ob_f}\\
&R^s_{k,c} \geq 0,~\forall k,\label{ob_g}\\
&\mathbf{F}_{n,m}=\left\{e^{j\theta}|\theta\in\left(0,2\pi\right]\right\},~\forall n,m, \label{ob_h} 
	\end{align}
\end{subequations}
where $\Gamma_{\theta_{\text{e}}}$,  $\Gamma_{r_{\text{e}}}$, and $P_{\text{th}}$ denote the given thresholds for angle estimation, distance estimation, and maximum transmit power, respectively. (\ref{ob_b}), (\ref{ob_c}), and (\ref{ob_d}) impose the requirements on sensing accuracy and transmit power while (\ref{ob_e}) guarantees that the common stream can act as artificial noise. In addition, (\ref{ob_f}) and (\ref{ob_g}) enforce the common secrecy rate allocation requirement, and (\ref{ob_h}) indicates the unit-modulus condition of analog beamfocuser $\mathbf{F}$.

Solving problem (\ref{linear_p}) entails three major technical challenges. First, the secrecy rate and the estimation error are non-convex and non-smooth, which hinders tractable optimization in the primal domain. Furthermore, solving the problem in the dual domain is also difficult due to an unknown duality gap. Second, analog beamfocuser $\mathbf{F}$ and digital beamfocuser $\mathbf{W}$ are tightly coupled and cannot be separated, making their joint optimization highly challenging. Third, unit-modulus constraint (\ref{ob_h}) further complicates the solution process. Consequently, no generic solution exists for problem (\ref{linear_p}), and obtaining the global optimum is practically intractable.

\section{Proposed algorithm}\label{Section III}
This section introduces auxiliary variables to reformulate the original problem (\ref{linear_p}) and subsequently develops a penalty-based BCD algorithm. It leverages WMMSE, quadratic transform, and Taylor expansion techniques to recast the communication rate, eavesdropping rate, and CRB constraints into more tractable forms. Furthermore, its convergence and computational complexity are discussed.
\subsection{Optimization problem reformulation}
The coupling between analog beamfocuser $\mathbf{F}$ and digital beamfocuser $\mathbf{W}$ in (\ref{linear_p}) makes direct optimization intractable. Notably, $\mathbf{F}$ and $\mathbf{W}$ always appear in product form, so we introduces a fully digital beamfocuser  $\mathbf{P}=\left[\mathbf{p}_0,\mathbf{p}_1,\dots, \mathbf{p}_K\right]\in\mathbb{C}^{N\times (K+1)}$ as an auxiliary variable, where $\mathbf{p}_k=\mathbf{Fw}_k$ for $\forall k\in\mathcal{K}_1=\{0,1,\dots,K\}$. By substituting $\mathbf{p}_k=\mathbf{Fw}_k$ into (\ref{power}), the average received power at the $k$-th communication user and the target can be reformulated as~\cite{7555358}
\begin{subequations}\label{power_2}
	\begin{align}
&T_{k,c}=\overbrace{{\left|\mathbf{h}^H_{k}\mathbf{p}_0\right|}^2}^{S_{k,c}}+\underbrace{\overbrace{{\left|\mathbf{h}^H_{k}\mathbf{p}_{k}\right|}^2}^{S_{{k,p}}}+\overbrace{\sum_{i=1,i\neq k}^{K}\left|\mathbf{h}^H_{k}\mathbf{p}_i\right|^2+\sigma^2_k}^{I_{{k,p}}}}_{I_{k,c}=T_{k,p}},
	\label{equ:power_legitimate_2}\\
&T_{\text{e},c}=\overbrace{{\left|\mathbf{g}^H_\text{e}\mathbf{p}_0\right|}^2}^{S_{\text{e},c}}+\underbrace{\overbrace{{\left|\mathbf{g}^H_\text{e}\mathbf{p}_{k}\right|}^2}^{S_{{\text{e},k}}}+\overbrace{\sum_{i=1,i\neq k}^{K}\left|\mathbf{g}^H_\text{e}\mathbf{p}_i\right|^2+\sigma^2_\text{e}}^{I_{{\text{e},k}}}}_{I_{\text{e},c}=T_{\text{e},k}}.
	\label{equ:power_eavesdropper_2}
	\end{align}
\end{subequations}
The above two identities are then incorporated into (\ref{Legiti_SINR}) and (\ref{Eaves_SINR}) to recalculate the legitimate and eavesdropping rates. Similarly, by substituting $\mathbf{p}_k=\mathbf{Fw}_k$ into (\ref{CRB}), the angle and distance estimation errors can be derived, as expressed in (\ref{Error_angle}) and (\ref{Error_distance}), with detailed derivations provided in Appendix~B.  
\begin{figure*}[!t]
\normalsize
\begin{align}
\text{CRB}\left(\theta_{\text{e}}\right)
= 
\frac{\sigma^2_{\text{e}}\text{Tr}\!\left(\tilde{\mathbf{ G}}\mathbf{PP}^H\tilde{\mathbf{ G}}^H\right)}
{2T\left|\tilde{\beta}_{\text{e}}\right|^2\!\left(
\text{Tr}\!\left(\dot{\mathbf{G}}_{\theta_{\text{e}}}\mathbf{PP}^H\dot{\mathbf{G}}^H_{\theta_{\text{e}}}\right)
\text{Tr}\!\left(\tilde{\mathbf{ G}}\mathbf{PP}^H\tilde{\mathbf{ G}}^H\right)
-\left|\text{Tr}\!\left(\tilde{\mathbf{ G}}\mathbf{PP}^H\dot{\mathbf{G}}^H_{\theta_{\text{e}}}\right)\right|^2
\right)}. \label{Error_angle}
\\[4pt]
\text{CRB}\left(r_{\text{e}}\right)
= 
\frac{\sigma^2_{\text{e}}\text{Tr}\!\left(\tilde{\mathbf{ G}}\mathbf{PP}^H\tilde{\mathbf{ G}}^H\right)}
{2T\left|\tilde{\beta}_{\text{e}}\right|^2\!\left(
\text{Tr}\!\left(\dot{\mathbf{G}}_{r_{\text{e}}}\mathbf{PP}^H\dot{\mathbf{G}}^H_{r_{\text{e}}}\right)
\text{Tr}\!\left(\tilde{\mathbf{ G}}\mathbf{PP}^H\tilde{\mathbf{ G}}^H\right)
-\left|\text{Tr}\!\left(\tilde{\mathbf{ G}}\mathbf{PP}^H\dot{\mathbf{G}}^H_{r_{\text{e}}}\right)\right|^2
\right)}. \label{Error_distance}
\end{align}
\hrule
\end{figure*}

The equality constraint $\mathbf{P}=\mathbf{FW}$ renders the optimization problem intractable. To address this challenge, the constraint is absorbed into the objective function as a penalty term. In addition, a non-negative auxiliary variable $R^s$ is introduced to eliminate the non-smoothness of the objective function. Based on these transformations, problem (\ref{linear_p}) can be reformulated into the augmented Lagrangian form as
\begin{subequations}\label{linear_p2}
	\begin{align}
&\max_{\mathbf{P},\mathbf{F},\mathbf{W},R^s_{k,c},R^s,R^s_{k,p} } R^s -\frac{1}{\rho}||\mathbf{P}-\mathbf{FW}||^2_F,\label{ob_a2}\\
	\text{s.t.}~
    &\text{CRB}\left(\theta_{\text{e}}\right)\leq\Gamma_{\theta_{\text{e}}},\label{ob_b2}\\
&\text{CRB}\left(r_{\text{e}}\right)\leq\Gamma_{r_{\text{e}}},\label{ob_c2}\\
&||\mathbf{P}||^2_F\leq P_{\text{th}},\label{ob_d2}\\
    &R^s\leq R^s_{k,c}+R^s_{k,p},~\forall k,\label{ob_e2}\\
    &R^s_{k,p} \leq R_{k,p}-R_{\text{e},k},~\forall k,\label{ob_f2}\\
    &\sum_{k=1}^{K}R^s_{k,c} \leq R_{k,c}-R_{\text{e},c},~\forall k,\label{ob_g2}\\
	&\mbox{ (\ref{ob_g}),~(\ref{ob_h})}, \label{ob_h2} 
	\end{align}
\end{subequations}
In (\ref{linear_p2}), the operator ${[\bullet]}^+$ is omitted without affecting optimality, which can be easily established through contradiction. In problem (\ref{linear_p2}), the coefficient $\rho>0$ denotes the penalty parameter. As $\rho$ decreases toward zero, the solution approaches the feasible point, where $\mathbf{P}=\mathbf{FW}$. However, selecting an excessively small $\rho$ at the early stage may cause the penalty term to dominate, thereby obscuring the secrecy rate maximization. To attack this issue, $\rho$ is initially set to a sufficiently large value to establish a stable starting point and is then progressively reduced until the equality constraint is fulfilled within a prescribed tolerance. This procedure naturally leads to a two-tier optimization framework, where the inner loop solves the problem with fixed $\rho$, and the outer loop adaptively adjusts $\rho$ to guarantee final feasibility.

Even with a fixed penalty parameter, problem~(\ref{linear_p2}) remains highly intractable. To address it, we partition the optimization variables into three disjoint blocks:
$
\mathcal{Q}=\{\mathbf{P}, R^s_{k,c}, R^s, R^s_{k,p}\},\; \mathbf{F},\; \text{and}\;  \mathbf{W}.
$
We then apply a BCD framework to decompose the original problem into three subproblems with significantly lower computational complexity. Specifically, in the fully digital beamfocusing stage, we develop an iterative algorithm to compute the legitimate and eavesdropper rates as well as the sensing CRBs. For the analog beamfocuser, we also derive an iterative update rule, whereas the digital beamfocuser admits a closed-form solution. These three stages are optimized sequentially and alternately until convergence. The detailed algorithms are presented in Sections \ref{Section III}.B-\ref{Section III}.E.

\subsection{Subproblem with respect to $\mathcal{Q}$}
Given fixed $\mathbf{F}$ and $\mathbf{W}$, the optimization problem for $\mathcal{Q}$ can be reformulated as
\begin{subequations}\label{linear_p3}
	\begin{align}
&\max_{\mathcal{Q}} R^s -\frac{1}{\rho}||\mathbf{P}-\mathbf{FW}||^2_F,\label{ob_a3}\\
	\text{s.t.}~
	&\mbox{ (\ref{ob_g}),~(\ref{ob_b2})~--~(\ref{ob_g2})}. \label{ob_b3} 
	\end{align}
\end{subequations}
Problem (\ref{linear_p3}) is non-convex due to fractional SINR, angle CRB, and distance CRB. In addition, the secrecy rate involves the difference of two logarithmic functions, aggravating the solution difficulty. 

To efficiently tackle problem (\ref{linear_p3}), we employ WMMSE, quadratic transform, and Taylor expansion techniques to recast the legitimate rate, eavesdropping rate, and sensing CRBs, respectively. Then, an iterative algorithm is developed to solve the resulting problem.

\subsubsection{Legitimate rate reformulation} For the $k$-th communication user, an equalizer $\omega_{k,c}$ is first applied to received signal $y_{k}(t)$ to recover the common message, yielding the estimate $\hat{x}_{k,c}=\omega_{k,c}y_k$, where time index $t$ is omitted for simplicity. Once the common stream is decoded and subtracted through SIC, another equalizer $\omega_{k,p}$ is employed to detect its private message, leading to $\hat{x}_{k}=\omega_{k,p}\left(y_k-\mathbf{h}^H_k\mathbf{p}_0x_0\right)$. The MSEs associated with decoding $x_0$ and $x_k$ are calculated as
\begin{subequations}\label{MMSE_error}
\begin{align}
\delta_{k,c}&=\mathbb{E}\left\{\left|\hat{x}_{k,c}-x_{k,c}\right|^2\right\}\notag\\ &=\left|\omega_{k,c}\right|^2T_{k,c}-2\text{Re}\left(\omega_{k,c}\mathbf{h}^H_k\mathbf{p}_0\right)+1,\\
\delta_{k,p}&=\mathbb{E}\left\{\left|\hat{x}_{k}-x_{k}\right|^2\right\}\notag\\&=\left|\omega_{k,p}\right|^2T_{k,p}-2\text{Re}\left(\omega_{k,p}\mathbf{h}^H_k\mathbf{p}_k\right)+1.
\end{align}
\end{subequations}
The minimum MSE (MMSE) equalizers are then calculated by solving $\frac{\partial \delta_{k,c}}{\partial \omega_{k,c}}=0$ and $\frac{\partial \delta_{k,p}}{\partial \omega_{k,p}}=0$, which are given by
\begin{subequations}
\begin{align}
\omega^{\text{MMSE}}_{k,c}=\mathbf{p}^H_0\mathbf{h}_kT^{-1}_{k,c},\\
\omega^{\text{MMSE}}_{k,p}=\mathbf{p}^H_k\mathbf{h}_kT^{-1}_{k,p}.
\label{Optimal_equalizer}
\end{align}
\end{subequations}
Substituting these optimum equalizers into (\ref{MMSE_error}), the MMSEs become
\begin{align}\label{MMSE_error2}
\delta^{\text{MMSE}}_{k,\psi} =\min_{\omega_{k,\psi}}\delta_{k,\psi}= &T^{-1}_{k,\psi}I_{k,\psi},~\forall \psi\in\{c,p\}.
\end{align}
Plugging (\ref{MMSE_error2}) back into (\ref{Legiti_SINR}), the corresponding SINR can be recast as 
$\gamma_{k,\psi}=1/\delta^{\text{MMSE}}_{k,\psi}-1$, from which the transmit rate rewrites $R_{k,\psi}=-\log\big(\delta^{\text{MMSE}}_{k,\psi}\big)$ for $\forall \psi\in\{c,p\}$.

The  augmented weighted MSEs (WMSEs) are then defined as 
\begin{align}\label{Relationship_2}
\beta_{k,\psi}&=\eta_{k,\psi}\delta_{k,\psi}-\log\left(\eta_{k,\psi}\right),~\forall \psi\in\{c,p\},
\end{align}
where $\eta_{k,\psi}$ denotes the positive weights. Consequently, the rate--WMMSE relationships are established as 
\begin{align}\label{Relationship}
\beta^{\text{MMSE}}_{k,\psi} &=\min_{\eta_{k,\psi},\omega_{k,\psi}}\beta_{k,\psi}= \tau-R_{k,\psi},~\forall \psi\in\{c,p\},
\end{align}
where $\tau=1/\ln 2 +\log(\ln 2)$. We observe that WMMSE is concave in each variable when the other two are fixed. Moreover, by checking the first optimality conditions, the optimal equalizers and weights are given by 
\begin{subequations}
\begin{align}\label{Optimal_WMMSE}
\omega^*_{k,\psi}&=\omega^{\text{MMSE}}_{k,\psi}, ~\forall \psi\in\{c,p\},\\
\eta^*_{k,\psi}&=\left(\delta^{\text{MMSE}}_{k,\psi}\ln 2\right)^{-1}, ~\forall \psi\in\{c,p\}.
\end{align}
\end{subequations}

{\bf{Remark 1:}} To ensure that constraints (\ref{ob_f2}) and (\ref{ob_g2}) define a convex feasible set while preserving solution feasibility to problem (\ref{linear_p3}), conservative approximations should be employed. In particular, strictly lower-bounded concave surrogates are constructed for legitimate rates $R_{k,c}$ and $R_{k,p}$, while strictly upper-bounded convex surrogates are imposed on eavesdropping rates $R_{\text{e},c}$ and $R_{\text{e},k}$. The WMMSE framework can reconstruct the legitimate rates as fixed equalizers and weights provide lower-bounded concave surrogates. However, it cannot provide upper-bounded convex approximations for the eavesdropping rates.

\subsubsection{Eavesdropping rate reformulation} To seek an alternative solution, we rewrite 
\begin{equation}
-R_{\text{e},c}=
\log\left(\frac{\sum^{K}_{i=1}\left|\mathbf{g}^H_\text{e}\mathbf{p}_i\right|^2+\sigma^2_{\text{e}}}{\sum^{K}_{i=0}\left|\mathbf{g}^H_\text{e}\mathbf{p}_i\right|^2+\sigma^2_{\text{e}}}\right),
\label{eavesdrop_negative}
\end{equation} which can be reformulated using the quadratic transform method~\cite{shen2018fractional}. However, this approach requires introducing $K+1$ vectors to decouple the fractional expression as the numerator contains $K+1$ terms (See~\cite{9721222} for details). Fortunately, we observe $\mathbf{g}^H_{\text{e}}\mathbf{g}_{\text{e}}=N|\beta_{\text{e}}|^2$, so the numerator in (\ref{eavesdrop_negative}) can be equivalently reformulated as 
\begin{equation}
\sum^{K}_{i=1}\left|\mathbf{g}^H_\text{e}\mathbf{p}_i\right|^2+\sigma^2_{\text{e}}=\left|\left|\mathbf{g}^H_\text{e}\mathbf{T}_0\right|\right|^2
\label{reconstruction}
\end{equation}
with
\begin{equation}
\mathbf{T}_0=\left[\frac{\sigma^2_{\text{e}}}{N|\beta_{\text{e}}|^2}\mathbf{g}_{\text{e}},\mathbf{p}_{1},\dots,\mathbf{p}_K\right].
\end{equation}
With this insight, we have 
\begin{align}
-R_{\text{e},c}=\log\left(\frac{\left|\mathbf{g}^H_\text{e}\mathbf{T}_0\right|^2}{T_{{\text{e}},c}}\right)
\overset{(a)}{=} \max_{\mathbf{x}_{\text{e},c}}f_{{\text{e}},c}\left(\mathbf{x}_{\text{e},c},\mathbf{P}\right)
\end{align}
where
\begin{align}
f_{{\text{e}},c}\left(\mathbf{x}_{\text{e},c},\mathbf{P}\right)= \log\left(2\text{Re}\left(\mathbf{x}^H_{\text{e},c}\mathbf{T}^H_0\mathbf{g}_\text{e}\right)-\mathbf{x}^H_{\text{e},c}T_{\text{e},c}\mathbf{x}_{\text{e},c}\right),\label{surrogate_common}
\end{align}
and $\mathbf{x}_{\text{e},c}\in\mathbb{C}^{N\times 1}$ is the introduced auxiliary variable. Equality $\overset{(a)}{=}$ follows from the quadratic transform method~\cite{shen2018fractional}. The optimal solution for $\mathbf{x}^*_{\text{e},c}$ is 
\begin{equation}
\mathbf{x}^*_{\text{e},c}=T_{\text{e},c}^{-1}\mathbf{T}^H_0\mathbf{g}_\text{e}.
\label{Optimal_common}
\end{equation}
which can be derived by solving $\frac{\partial f_{{\text{e}},c}\left(\mathbf{x}_{\text{e},c},\mathbf{P}\right)}{\partial \mathbf{x}_{\text{e},c}}=0$. Similarly, eavesdropping rate $-R_{e,k}$ for $\forall k$ can be reformulated as
\begin{align}
-R_{e,k}=\log\left(\frac{\left|\mathbf{g}^H_{\text{e}}\mathbf{T}_k\right|^2}{T_{{\text{e}},c}}\right)=\max_{\mathbf{x}_{\text{e},k}}f_{{\text{e}},k}\left(\mathbf{x}_{\text{e},k},\mathbf{P}\right),
\end{align}
where 
\begin{align}
f_{{\text{e}},k}\left(\mathbf{x}_{\text{e},k},\mathbf{P}\right)=\log\left(2\text{Re}\left(\mathbf{x}^H_{\text{e},k}\mathbf{T}^H_k\mathbf{g}_\text{e}\right)-\mathbf{x}^H_{\text{e},k}T_{\text{e},c}\mathbf{x}_{\text{e},k}\right),\label{surrogate_private}
\end{align}
and
\begin{equation}
\mathbf{T}_k=\left[\mathbf{p}_0,\dots,\mathbf{p}_{k-1},\frac{\sigma^2_{\text{e}}}{N|\beta_{\text{e}}|^2}\mathbf{g}_{\text{e}},\mathbf{p}_{k+1},\dots,\mathbf{p}_K\right].
\end{equation}
Optimal auxiliary variable $\mathbf{x}_{\text{e},k}$ is given by
\begin{equation}
\mathbf{x}^*_{\text{e},k}=T_{\text{e},c}^{-1}\mathbf{T}^H_k\mathbf{g}_\text{e}.
\label{Optimal_private}
\end{equation}

\subsubsection{Angle and distance CRB reformulation} Based on (\ref{Error_angle}), the angle and distance CRB constraints (\ref{ob_b2}) and (\ref{ob_c2}) can be reformulated as
\begin{align}
\text{Tr}\left(\dot{\mathbf{G}}_{s}\mathbf{PP}^H\dot{\mathbf{G}}^H_{s}\right)-\frac{\left|\text{Tr}\left(\tilde{\mathbf{ G}}\mathbf{PP}^H\dot{\mathbf{G}}^H_{s}\right)\right|^2}{\text{Tr}\left(\tilde{\mathbf{ G}}\mathbf{PP}^H\tilde{\mathbf{ G}}^H\right)}\geq \tilde \Gamma_{s},
\label{CRB constraint}
\end{align}
where $\tilde \Gamma_{\theta_{\text{e}}}=\left(\sigma^2_{\text{e}}\right)\left(2T\left|\tilde{\beta}_{\text{e}}\right|^2\Gamma_{s}\right)^{-1}$ and $s\in\{\theta_{\text{e}},r_{\text{e}}\}$. Introducing non-negative auxiliary variables $\mu_s$ and $\alpha_s$, (\ref{CRB constraint}) can be equivalently decomposed into three constraints:
\begin{subequations}
\begin{align}
\left|\text{Tr}\left(\tilde{\mathbf{ G}}\mathbf{PP}^H\dot{\mathbf{G}}^H_{s}\right)\right|&\leq\mu_s,\label{crb_a}\\
\text{Tr}\left(\tilde{\mathbf{ G}}\mathbf{PP}^H\tilde{\mathbf{ G}}^H\right)&\geq\alpha_s,\label{crb_b}\\
\text{Tr}\left(\dot{\mathbf{G}}_{s}\mathbf{PP}^H\dot{\mathbf{G}}^H_{s}\right)&\geq \frac{\mu^2_s}{\alpha_s}+\tilde \Gamma_{s}.\label{crb_c}
\end{align}
\end{subequations} 
The left-hand side of (\ref{crb_a}) is non-convex with respect to $\mathbf{P}$ since $\dot{\mathbf{G}}^H_{s}\tilde{\mathbf{ G}}$ is generally neither positive semi-definite nor negative semi-definite. Moreover, the combination of modulus, trace, and matrix multiplication complicates its reformulation. To address this, we introduce Lemma~1, which is proved in Appendix~C

{\bf{Lemma 1 :}} Let $\mathbf{A}\in\mathbb{C}^{N\times N}$ be an arbitrary constant matrix and $\mathbf{X}\in\mathbb{C}^{N\times N}$ be a Hermitian positive semi-definite matrix, i.e., $\mathbf{X}\succeq\mathbf{0}$. Define the Hermitian part of $\mathbf{A}$ as $\mathbf{S}\triangleq \frac{\mathbf{A}+\mathbf{A}^H}{2}$. Let the eigen decomposition of $\mathbf{S}$ be $\mathbf{S}=\mathbf{U}\mathbf{D}\mathbf{U}^H$, where $\mathbf{U}$ is unitary and $\mathbf{D}=\text{diag}(d_1,\dots,d_N)$ is real diagonal.
Then, the following equality holds,
\begin{align}
\text{Re}\left(\text{Tr}(\mathbf{A}\mathbf{X})\right)
= \text{Tr}\left(\mathbf{U}\mathbf{D}_+\mathbf{U}^H\mathbf{X}\right)
- \text{Tr}\left(\mathbf{U}\mathbf{D}_-\mathbf{U}^H\mathbf{X}\right).
\label{DC}
\end{align}
where
\begin{subequations}
\begin{align}
\mathbf{D}_+ &\triangleq \text{diag}(\max(d_i,0)),\\
\mathbf{D}_- &\triangleq \text{diag}(\max(-d_i,0)).
\end{align}
\end{subequations} 

To proceed, we rewrite
\begin{align}
&\left|\text{Tr}\left(\tilde{\mathbf{ G}}\mathbf{PP}^H\dot{\mathbf{G}}^H_{s}\right)\right|\notag\\=&\left|\text{Re}\left(\text{Tr}\left(\mathbf{A}_{1}\mathbf{PP}^H\right)\right)+j\text{Re}\left(\text{Tr}\left(\mathbf{A}_{2}\mathbf{PP}^H\right)\right)\right|,
\end{align}
where $\mathbf{A}_1=\dot{\mathbf{G}}^H_{s}\tilde{\mathbf{ G}}$ and $\mathbf{A}_2=-j\dot{\mathbf{G}}^H_{s}\tilde{\mathbf{G}}$. According to Lemma~1, $\text{Re}\big(\text{Tr}(\mathbf{A}_{x}\mathbf{PP}^H)\big)$ for $x\in\{1,2\}$ can be expressed as a difference of two convex quadratic forms. Specifically, let $\mathbf{p}=\mathrm{vec}(\mathbf{P})$, we have
\begin{equation}
g_x(\mathbf{p}) \triangleq \text{Re}\big(\text{Tr}(\mathbf{A}_{x}\mathbf{PP}^H)\big)
= \mathbf{p}^H\mathbf{M}_{x,+}\mathbf{p} - \mathbf{p}^H\mathbf{M}_{x,-}\mathbf{p},
\label{eq:g-original}
\end{equation}
where $\mathbf{M}_{x,\pm} \triangleq \mathbf{I}_{K+1}\otimes(\mathbf{U}_x\mathbf{D}_{x,\pm}\mathbf{U}^H_x)$, \(\mathbf{D}_{x,\pm}\) are the positive and negative parts of the Hermitian part \(\mathbf{S}_x=(\mathbf{A}_x+\mathbf{A}^H_x)/2\), and \(\mathbf{U}_x\mathbf{D}_x\mathbf{U}^H_x\) is the eigen decomposition of \(\mathbf{S}_x\). Since $\mathbf{M}_{x,-}\succeq 0$, the second term in \eqref{eq:g-original} is convex. Using first-order Taylor expansion at $\tilde{\mathbf{p}}$, $g_x(\mathbf{p})$  can be lower bounded as
\begin{align}
\hat g_x(\mathbf{p};\tilde{\mathbf{p}})
= \mathbf{p}^H\mathbf{M}_{x,+}\mathbf{p}
- 2\text{Re}\big(\tilde{\mathbf{p}}^H\mathbf{M}_{x,-}\mathbf{p}\big)
+ \tilde{\mathbf{p}}^H\mathbf{M}_{x,-}\tilde{\mathbf{p}}.
\label{eq:surrogate}
\end{align}
As a result, constraint (\ref{crb_a}) can be conservatively approximated as a second-order cone (SOC) constraint
\begin{align}\label{CRB_recons_1}
\sqrt{\hat g^2_1(\mathbf{p};\tilde{\mathbf{p}}) +g^2_2(\mathbf{p};\tilde{\mathbf{p}})}\leq \mu_s, \forall s\in\{\theta_{\text{e}},r_\text{e}\},
\end{align}
which remains a convex set. Similarly, applying the first-order Taylor expansion at $\tilde{\mathbf{p}}$, (\ref{crb_b}) and (\ref{crb_c}) can be approximated as
\begin{subequations}\label{CRB_recons_2}
\begin{align}
&2\text{Re}\left(\tilde{\mathbf{p}}^H\bar{\mathbf{G}}\mathbf{p}\right)-\tilde{\mathbf{p}}^H\bar{\mathbf{G}}\tilde{\mathbf{p}}\geq \alpha_{s},~\forall s\in\{\theta_{\text{e}},r_\text{e}\},\\
&2\text{Re}\left(\tilde{\mathbf{p}}^H\ddot{\mathbf{G}}_s\mathbf{p}\right)-\tilde{\mathbf{p}}^H\ddot{\mathbf{G}}_s\tilde{\mathbf{p}}\geq \frac{\mu^2_s}{\alpha_s}+\tilde \Gamma_{s},~\forall s\in\{\theta_{\text{e}},r_\text{e}\}.
\end{align}
\end{subequations} 
where $\ddot{\mathbf{G}}_s=\mathbf{I}_{K+1}\otimes\left(\dot{\mathbf{G}}^H_{s}\dot{\mathbf{G}}_{s}\right)$ and $\bar{\mathbf{G}}= \mathbf{I}_{K+1}\otimes\left(\tilde{\mathbf{G}}^H\tilde{\mathbf{G}}\right)$.

Based on the above reformulations, problem (\ref{linear_p3}) can be recast to
\begin{subequations}\label{linear_p4}
	\begin{align}
&\max_{\mathcal{Q},\tilde{\mathcal{Q}}} R^s -\frac{1}{\rho}||\mathbf{P}-\mathbf{FW}||^2_F,\label{ob_a4}\\
	\text{s.t.}~&R^s_{k,p} \leq \tau -\min_{\eta_{k,p},\omega_{k,p}}\beta_{k,p} +\max_{\mathbf{x}_{\text{e},k}}f_{{\text{e}},k}\left(\mathbf{x}_{\text{e},k},\mathbf{P}\right),~\forall k,\label{ob_b4}\\
    &\sum_{k=1}^{K}R^s_{k,c} \leq \tau-\min_{\eta_{k,c},\omega_{k,c}}\beta_{k,c}+\max_{\mathbf{x}_{\text{e},c}}f_{{\text{e}},c}\left(\mathbf{x}_{\text{e},c},\mathbf{P}\right),~\forall k,\label{ob_c4}\\
	&\mbox{ (\ref{ob_g}),~(\ref{ob_d2}),~(\ref{ob_e2}),~(\ref{CRB_recons_1}),~(\ref{CRB_recons_2})}. \label{ob_d4} 
	\end{align}
\end{subequations}
where $\tilde{\mathcal{Q}}=\left\{\eta_{k,p},\omega_{k,p},\eta_{k,c},\omega_{k,c},\mathbf{x}_{\text{e},k},\mathbf{x}_{\text{e},c}\right\}$. It can be observed that problem (\ref{linear_p4}) becomes convex for a fixed auxiliary variable set $\tilde{\mathcal{Q}}$, while the optimal $\tilde{\mathcal{Q}}$ can be derived in closed-form via (\ref{Optimal_WMMSE}), (\ref{Optimal_common}), and (\ref{Optimal_private}). Motivated by this property, we alternately optimize $\mathcal{Q}$ and $\tilde{\mathcal{Q}}$ in an iterative manner. The overall procedure is summarized in Algorithm~\ref{Alg.1}.
\begin{algorithm}[t]
	\caption{Iterative algorithm for solving (\ref{linear_p2})}
	\begin{algorithmic}[1]\label{Alg.1}
		\STATE Initialize a feasible $\mathbf{P}$ and set a predefined threshold $\epsilon_1$. 
		\REPEAT
        \STATE Update $\tilde{\mathbf{P}}=\mathbf{P}$.
        \STATE Update optimal     
        $\tilde{\mathcal{Q}}$ based on (\ref{Optimal_WMMSE}), (\ref{Optimal_common}), and (\ref{Optimal_private}).
		\STATE  Solving problem (\ref{linear_p4}) to obtain updated $\mathbf{P}$.
		\UNTIL{the increment of the objective value in (\ref{linear_p4}) is smaller than $\epsilon_1$.}	
		\STATE Return the optimized $\mathbf{P}$.
	\end{algorithmic}
\end{algorithm}

\subsection{Subproblem with respect to $\mathbf{F}$} 
When $\mathbf{P}$ and $\mathbf{W}$ are fixed, problem (\ref{linear_p2}) can be rewritten as
\begin{subequations}\label{linear_p10_new}
	\begin{align}
&\min_{\mathbf{F}}~ \mathrm{Tr}\left(\mathbf{F}^H\mathbf{F}\mathbf{Y}\right)-2\text{Re}\left(\mathrm{Tr}\left(\mathbf{F}^H\mathbf{Z}\right)\right), \label{ob_a10_new}\\
\text{s.t.}~~
&\mathbf{F}_{n,m}=\left\{e^{j\theta}|\theta\in\left(0,2\pi\right]\right\},~\forall n\in\mathcal{N},~\forall m\in\mathcal{M}, \label{ob_b10_new}
	\end{align}
\end{subequations}
with $\mathbf{Y}=\mathbf{W}\mathbf{W}^H$ and $\mathbf{Z}=\mathbf{P}\mathbf{W}^H$. Since the unit-modulus constraint in (\ref{ob_b10_new}) acts on each entry individually, the problem is separable across elements of $\mathbf{F}$. Thus, optimizing a single coefficient $\mathbf{F}_{n,m}$ amounts to solving
\begin{subequations}\label{linear_p11_new}
	\begin{align}
&\min_{\mathbf{F}_{n,m}}~ \phi_{n,m}|\mathbf{F}_{n,m}|^2 - 2\text{Re}\left(\chi_{n,m}\mathbf{F}_{n,m}\right), \label{ob_a11_new}\\
\text{s.t.}~~
&\mathbf{F}_{n,m}=\left\{e^{j\theta}|\theta\in\left(0,2\pi\right]\right\}, \label{ob_b11_new}
	\end{align}
\end{subequations}
where $\phi_{n,m}$ and $\chi_{n,m}$ are constants dependent on all other entries $\mathbf{F}$. Enforcing the unit-modulus constraint directly yields
\begin{equation}
\mathbf{F}_{n,m}^* = e^{-j\angle{\chi_{n,m}}}.
\label{Optimal_analog_new}
\end{equation} 
Although $\chi_{n,m}$ is not explicitly available, we note that the objectives in (\ref{linear_p10_new}) and (\ref{linear_p11_new}) share identical gradients with respect to $\mathbf{F}_{n,m}$. Therefore, we obtain the relation
\begin{equation}
\mathbf{X}_{n,m}-\mathbf{Z}_{n,m} = \phi_{n,m}\tilde{\mathbf{F}}_{n,m}-\chi_{n,m},
\label{solution_2}
\end{equation}
where $\mathbf{X}=\tilde{\mathbf{F}}\mathbf{Y}$ and $\tilde{\mathbf{F}}$ denotes the analog beamfocuser from the previous iteration. Expanding $\tilde{\mathbf{F}}\mathbf{Y}$ further gives
\begin{align}
\phi_{n,m}\tilde{\mathbf{F}}_{n,m}=\tilde{\mathbf{F}}_{n,m}\mathbf{Y}_{m,m}
\label{solution}
\end{align}
Combining (\ref{solution_2}) and (\ref{solution}), we find 
\begin{equation}
\chi_{n,m}=\mathbf{Z}_{n,m}-\mathbf{X}_{n,m}+\tilde{\mathbf{F}}_{n,m}\mathbf{Y}_{m,m}.
\label{Optimal_F_new}
\end{equation} 
This yields the optimal solution for the analog beamfocusor. 
\subsection{Subproblem with respect to $\mathbf{W}$} 
Digital beamfocuser $\mathbf{W}$ appears only in the final quadratic term of the objective function. Hence, fixing $\mathbf{P}$ and $\mathbf{F}$ simplifies problem (\ref{linear_p3}) to $\min_{\mathbf{W}} \|\|\mathbf{P}-\mathbf{F}\mathbf{W}\|\|_F^2$,
which is unconstrained and convex in $\mathbf{W}$. Applying the first-order optimality condition gives the closed-form solution
\begin{align}\label{linear_p6_new}
\mathbf{W}^* = \left(\mathbf{F}^H\mathbf{F}\right)^{-1}\mathbf{F}^H\mathbf{P}.
\end{align}

\subsection{Overall algorithm}
Leveraging the block-wise solutions, the complete penalty-based BCD procedure is summarized in Algorithm~\ref{Alg.2}. We next provide an analysis of its convergence and computational complexity.
\begin{itemize}
\item  \emph{Convergence}: With any feasible initialization, both the inner and outer loops of Algorithm~\ref{Alg.2} generate non-decreasing objective values. In particular, each update of $\mathbf{P}$, $\mathbf{F}$, and $\mathbf{W}$ either improves or maintains the secrecy rate under a fixed penalty factor $\rho$. Since the secrecy rate is naturally upper-bounded, the inner loop is guaranteed to converge within finite iterations. Moreover, the outer loop is based on a penalty method, which has been established to converge to a stationary point \cite{9120361}. Hence, the proposed penalty-based BCD algorithm can be regarded as convergent.

\item \emph{Complexity}: The main computational cost comes from three steps: (i) updating $\mathbf{P}$ via Algorithm~\ref{Alg.1}, where the complexity scales as $\mathcal{O}\left(\delta_1 (N(K+1)+2K)^{3.5}\right)$ with $\delta_1$ denoting the iteration count of Algorithm~\ref{Alg.1}; (ii) updating $\mathbf{F}$, which requires complexity on the order of $\mathcal{O}(NM(K+1)^2)$; (iii) updating $\mathbf{W}$, which involves matrix multiplications of order $\mathcal{O}(NM\max(M,K+1))$. 

\end{itemize}

\begin{algorithm}[t]
	\caption{Penalty-based BCD algorithm for solving (\ref{linear_p3})}
	\begin{algorithmic}[1]\label{Alg.2}
		\STATE Initialize $\mathbf{F}$ and $\mathbf{W}$ and set a predefined threshold $\epsilon_2$.
        \REPEAT
		\REPEAT
		\STATE  Update $\mathbf{P}$ using Algorithm~\ref{Alg.1}.
        \STATE Update $\mathbf{F}$ according to using (\ref{Optimal_analog_new}).
        \STATE Update $\mathbf{W}$ according to (\ref{linear_p6_new}).
		\UNTIL { the improvement of the objective value becomes negligible. }
        \STATE Update penalty factor $\rho=\alpha  \rho$.
        \UNTIL{the penalty term falls below $\epsilon_2$. }
		\STATE Return the optimized secrecy rate $R^s$.
	\end{algorithmic}
\end{algorithm}

\section{Simulation results}\label{Section IV}
This section numerically evaluates the performance of the proposed secure transmit scheme. The default simulation setup is described as follows. The BS has $N=128$ transmit antennas, $M=64$ receive antennas, $L=8$ RF chains, operating at a carrier frequency of $f_c=30$~GHz. All antennas are arranged with half-wavelength spacing. $K=4$ legitimate users and one target are randomly distributed within a distance range of $[10,20]$ meters and an angular sector of $[0,\tfrac{\pi}{2}]$. The overall time slot is set to $T=1000$. The maximum transmit power is limited to $P_{\text{th}}=20$~dBm while the noise variance for both the legitimate users and the target is $\sigma^2_k=\sigma^2_{\text{e}}=-84$~dBm. For the proposed algorithm, the penalty factor is initialized at $\rho=10^2$ and updated with a reduction coefficient $\alpha=0.5$.

Our proposed secure transmit scheme is labeled as {\bf{RSMA-HB}}. Its performance is evaluated over $100$ independent spherical-wave channel realizations and compared against the following four benchmarks:
\begin{itemize}
\item {\bf{RSMA-FD}}: Each antenna is equipped with a dedicated RF chain. The formulated non-convex optimization problem can be solved using Algorithm~\ref{Alg.1}, serving as an upper bound for our scheme.
\item {\bf{RSMA-SC}}: This benchmark maximizes the minimum secrecy rate without considering sensing CRB constraints. It provides a reference to quantify the secrecy performance loss incurred when enforcing sensing accuracy requirements.
\item {\bf{SDMA-HB}}: This benchmark is implemented by disabling the common stream (i.e., $\mathbf{w}_0=\mathbf{0}$). It relies exclusively on near-field beamfocusing to prevent information leakage while simultaneously performing target sensing.

\item {\bf{RSMA-Far}}: This one is based on the conventional far-field plane-wave channel model, where the array response vector for device $i$ is given by
\begin{align}
\left[\mathbf{a}\left(\theta_k\right)\right]_n=e^{j\frac{2\pi}{\lambda}nd\sin\theta_k}.
\label{Far-Channel}
\end{align}
with $\forall i\in\left\{1,\dots,K,\text{e}\right\}$. However, the sensing channel continues to adopt the spherical-wave model to ensure joint distance–angle estimation.
\end{itemize}

\begin{figure}[tbp]
	\centering
	\includegraphics[scale=0.6]{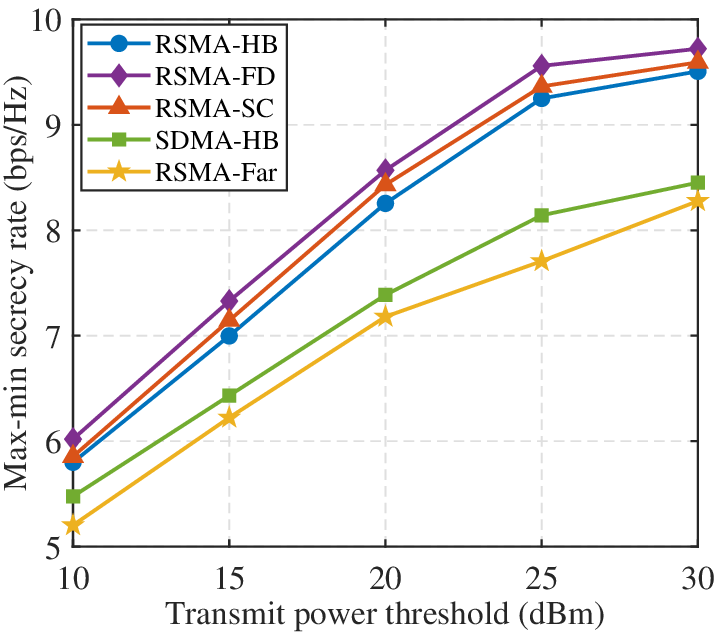}
	\caption{Max-min secrecy rate versus transmit power threshold.}
	\label{power}
\end{figure}
Fig.~\ref{power} illustrates the variation of the max–min secrecy rate with respect to the transmit power threshold. As the maximum transmit power increases, all schemes exhibit a monotonically rising secrecy rate. Several key insights can be observed. First, our proposed scheme nearly matches the performance of the fully digital architecture while using only eight RF chains, demonstrating its excellent hardware efficiency. Second, compared with RSMA-SC, which neglects the sensing CRB constraints, our design incurs only a minor secrecy rate loss, indicating that enforcing high-accuracy sensing introduces negligible degradation. This confirms the effectiveness of the common stream in jointly supporting communication, sensing, and security.  Third, the proposed scheme consistently outperforms SDMA across all transmit power levels, as SDMA relies solely on spatial beam separation and fails to suppress residual multi-user interference. In contrast, RSMA adaptively allocates power to the common stream, which mitigates partial interference and enhances secrecy by acting as artificial noise. Finally, the proposed near-field design delivers noticeable gains over the far-field plane-wave model, since spherical-wavefront beamfocusing confines energy within a localized region, effectively reducing signal leakage.

\begin{figure}[tbp]
	\centering
	\includegraphics[scale=0.6]{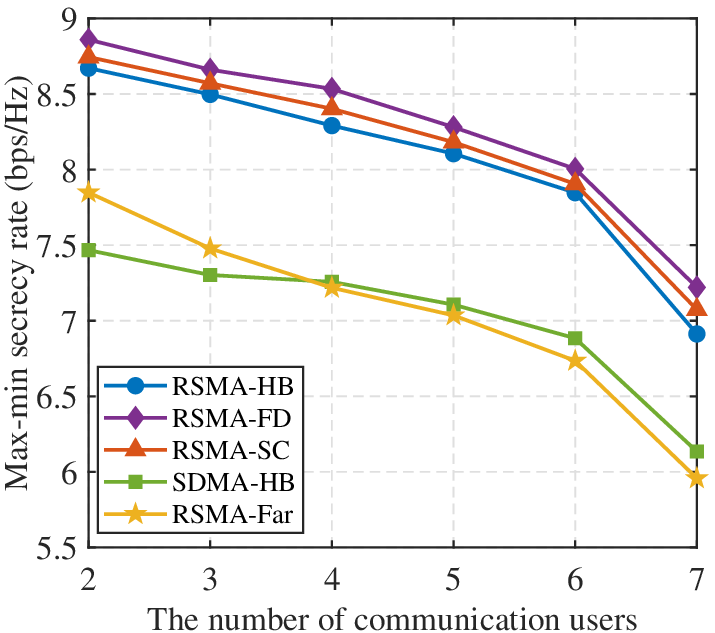}
	\caption{Max-min secrecy rate versus the number of users.}
	\label{user}
\end{figure}
Fig.~\ref{user} presents the max–min secrecy rate versus the number of communication users. As the user count increases, all schemes experience a gradual decline in the minimum secrecy rate due to aggravated multi-user interference. Nevertheless, the proposed scheme always surpasses both the SDMA-based and far-field counterparts, demonstrating strong robustness across varying user densities. This superiority arises from the combination of RSMA’s adaptive interference
management and the beamfocusing gain provided by spherical waves. A notable performance crossover is also observed: far-field RSMA degrades rapidly and is eventually overtaken by near-field SDMA when $K\geq 4$. This behavior is attributed to the common stream in RSMA, whose achievable rate is limited by the weakest user and shared by all users. As the number of users grows, the benefit of the common stream diminishes. In contrast, near-field SDMA progressively leverages spatial beamfocusing to mitigate multi-user interference. Consequently, when many users are scheduled, the near-field focusing gain becomes dominant, yielding superior secrecy performance.

\begin{figure}[tbp]
	\centering
	\includegraphics[scale=0.6]{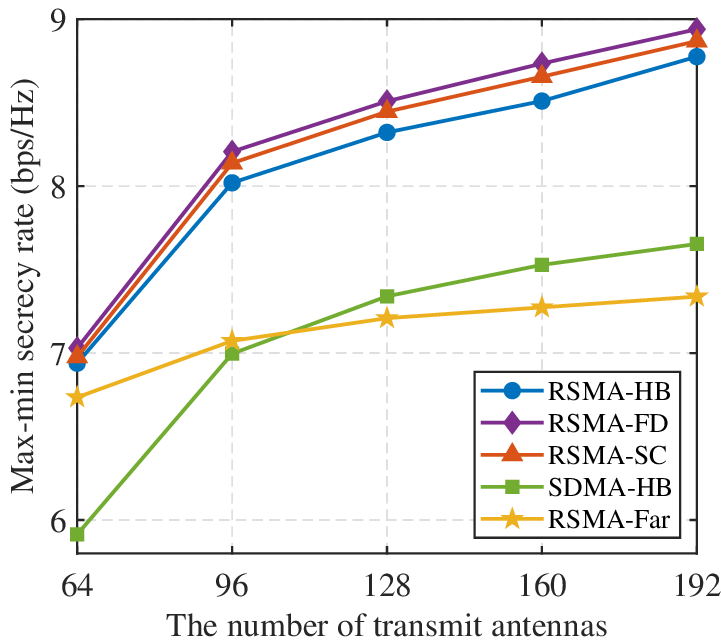}
	\caption{Max-min rate versus the number of transmit antennas.}
	\label{antenna}
\end{figure}
Fig.~\ref{antenna} shows the max–min secrecy rate as a function of the number of transmit antennas. As the array aperture expands, all schemes benefit from improved beamfocusing gain and high spatial resolution. Compared with the conventional SDMA scheme, the RSMA-enhanced design achieves an additional secrecy rate gain of approximately $1.2$ bps/Hz secrecy rate gain, underscoring the effectiveness of the common stream in mitigating interference and suppressing eavesdropping. Consistent with the trend observed in Fig.~\ref{user}, the near-field SDMA scheme outperforms its far-field RSMA counterpart when $N\geq 128$. Moreover, the superiority of the near-field configuration becomes increasingly pronounced with larger antenna arrays. These two phenomena highlight the superior signal enhancement and energy leakage suppression capabilities enabled by near-field beamfocusing.

\begin{figure}[tbp]
	\centering
	\includegraphics[scale=0.6]{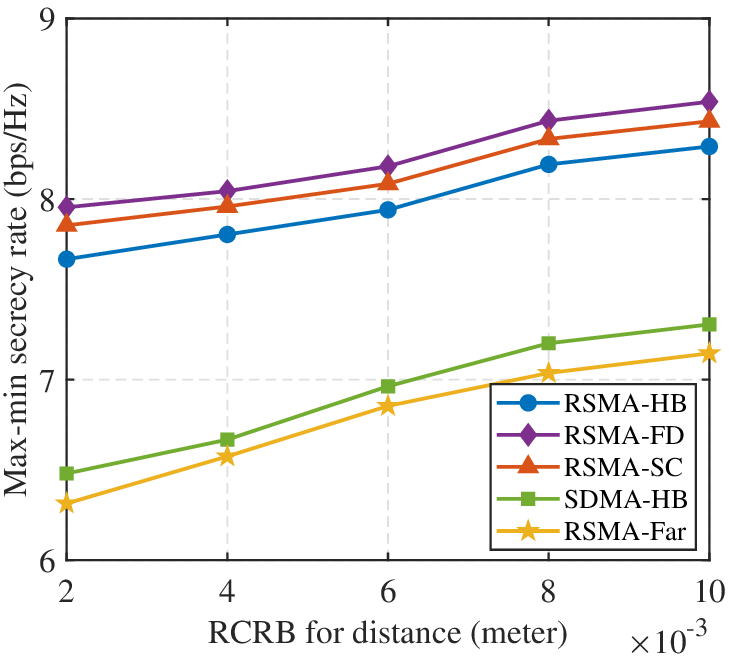}
	\caption{Max-min rate versus RCRB for distance.}
	\label{distance}
\end{figure}
Fig.~\ref{distance} depicts the max-min secrecy rate versus the distance RCRB requirement. Across all sensing accuracy requirements, our proposed scheme exhibits only a marginal performance loss compared with the RSMA-SC benchmark, indicating that incorporating high-precision sensing imposes negligible degradation on secure communications. The robustness stems from the common stream's ability to effectively serve triple functions. In contrast, both the SDMA-based and far-field schemes undergo severe secrecy degradation under tight RCRB requirements. The SDMA approach lacks flexibility in jointly allocating power between communication and sensing, whereas the far-field model fails to exploit near-field beamfocusing, resulting in reduced sensing efficiency and increased information leakage.

\begin{figure}[tbp]
	\centering
	\includegraphics[scale=0.6]{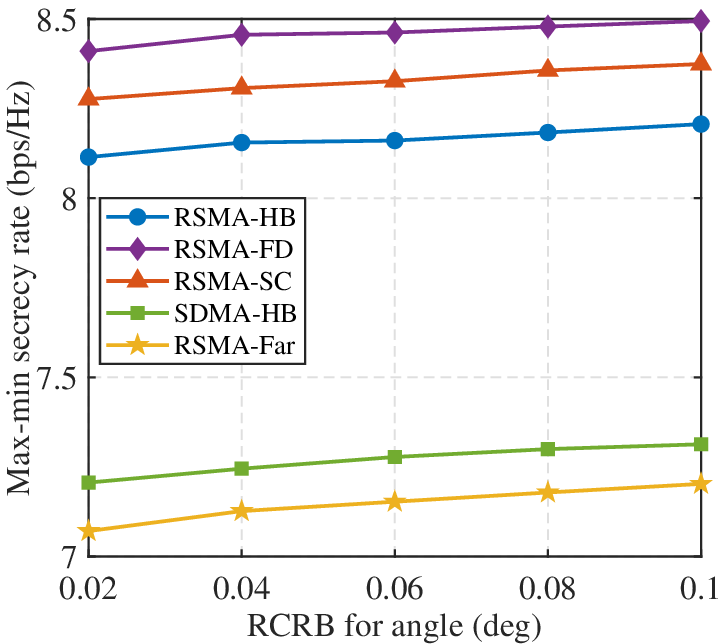}
	\caption{Max-min rate versus RCRB for angle.}
	\label{angle}
\end{figure}
Fig.~\ref{angle} depicts the max–min secrecy rate versus the RCRB requirement for angle estimation. It can be observed that all schemes remain almost stable across different angular sensing requirements, indicating that such constraints are readily met. Compared to the fully digital beamfocuser, our hybrid beamfocusing architecture incurs about a 4\% performance loss while reducing the number of RF chains by a factor of $16$. Moreover, the proposed scheme enables high-precision near-field sensing with minimal secrecy degradation and surpasses both conventional SDMA and far-field schemes across the considered sensing accuracy range.

\section{Conclusion}\label{Section V}
The HAD architecture in near-field ISAC reduces beamfocusing precision and causes energy leakage, which intensifies inter-user interference and elevates eavesdropping risks. To address these challenges, this paper proposes an RSMA-enhanced secure transmit design for near-field ISAC in which the common stream is strategically structured to serve three roles, namely interference reduction, sensing sequence, and artificial noise. We derive the angle and distance CRBs and use them to formulate a minimum-secrecy-rate maximization problem under these constraints. To handle the resulting non-convexity, we develop a penalty-based BCD algorithm that incorporates WMMSE, quadratic transforms, and Taylor expansions. Simulation results show that the proposed RSMA-enhanced scheme closely approaches the performance of fully digital beamfocusing while requiring far fewer RF chains. It delivers substantially stronger secrecy guarantees than conventional beamfocusing-only and far-field security schemes, and it achieves high-accuracy near-field sensing with minimal degradation to communication performance. 

Our work provides valuable insights that inspire several promising research directions. First, accurately estimating channel information poses greater challenges in near-field than in far-field scenarios. This motivates the exploration of RSMA's robustness to compensate for imperfect channel information. Second, given the mobility of targets/eavesdroppers in practical systems, another promising direction is to leverage ISAC's sensing capabilities to devise secure transmit schemes.

\appendices
\section{Derivation of FIM matrix}
According to~\cite{4359542}, the element at the $i$-th row and the $j$-th column of $\mathbf{J}_{\bm{\xi}}$ can be calculated by 
\begin{align}
\left[\mathbf{J}_{\bm{\xi}}\right]_{i,j}=& 2\text{Re}\left\{\frac{\partial \mathbf{u}^H}{\partial \bm{\xi}_i}\mathbf{v}^{-1}\frac{\partial \mathbf{u}}{\partial \bm{\xi}_j}\right\} + \text{Tr}\left(\mathbf{v}^{-1}\frac{\partial \mathbf{v}^H}{\partial \bm{\xi}_i}\mathbf{v}^{-1}\frac{\partial \mathbf{v}}{\partial \bm{\xi}_j}\right) \notag\\=& \frac{2}{\sigma^2_{\text{e}}}\text{Re}\left\{\frac{\partial \mathbf{u}^H}{\partial \bm{\xi}_i}\frac{\partial \mathbf{u}}{\partial \bm{\xi}_j}\right\},
\end{align}
where $\bm{\xi}_i$ denotes the $i$-th element of $\bm{\xi}$. Defining $\tilde{\mathbf{G}}=\mathbf{b}\left(r_\text{e},\theta_\text{e}\right)\mathbf{a}^T\left(r_\text{e},\theta_\text{e}\right)$, $\dot{\mathbf{G}}_{r_\text{e}} = \frac{\partial \tilde{\mathbf{G}}}{\partial r_{\text{e}}}$, and $\dot{\mathbf{G}}_{\theta_\text{e}} = \frac{\partial \tilde{\mathbf{G}}}{\partial \theta_{\text{e}}}$, we have
\begin{align}\label{partial_derivative}
&\frac{\partial \mathbf{u}}{\partial {r_{\text{e}}}} = \tilde{\beta}_{\text{e}} \text{vec}\left(\dot{\mathbf{G}}_{r}\mathbf{X}\right),\quad \frac{\partial \mathbf{u}}{\partial {\theta_{\text{e}}}} = \tilde{\beta}_{\text{e}} \text{vec}\left(\dot{\mathbf{G}}_{\theta}\mathbf{X}\right),\notag\\
&\frac{\partial \mathbf{u}}{\partial \tilde{\beta}^{\text{R}}_{\text{e}}} = \text{vec}\left(\tilde{\mathbf{G}}\mathbf{X}\right),\quad\frac{\partial \mathbf{u}}{\partial \tilde{\beta}^{\text{I}}_{\text{e}}} =j \text{vec}\left(\tilde{\mathbf{G}}\mathbf{X}\right).
\end{align}

The FIM matrix can be expressed as
\begin{align}\label{FIM_2}
\mathbf{J}_{\bm{\xi}}&=\frac{2T}{\sigma^2_0}\left[\begin{array}{cc|cc}
J_{\theta_{\text{e}}\theta_{\text{e}}} & J_{\theta_{\text{e}}r_{\text{e}}} & J_{\theta_{\text{e}}\tilde{\beta}^{\text{R}}_{\text{e}}} & J_{\theta_{\text{e}}\tilde{\beta}^{\text{I}}_{\text{e}}}\\
J_{\theta_{\text{e}}r_{\text{e}}}&J_{r_{\text{e}}r_{\text{e}}}&J_{r_{\text{e}}\tilde{\beta}^{\text{R}}_{\text{e}}}&J_{r_{\text{e}}\tilde{\beta}^{\text{I}}_{\text{e}}}\\
\hline 
J_{\theta_{\text{e}}\tilde{\beta}^{\text{R}}_{\text{e}}}&J_{r_{\text{e}}\tilde{\beta}^{\text{R}}_{\text{e}}}&J_{\tilde{\beta}^{\text{R}}_{\text{e}}\tilde{\beta}^{\text{R}}_{\text{e}}}&0\\
J_{\theta_{\text{e}}\tilde{\beta}^{\text{I}}_{\text{e}}}&J_{r_{\text{e}}\tilde{\beta}^{\text{I}}_{\text{e}}}&0&J_{\tilde{\beta}^{\text{I}}_{\text{e}}\tilde{\beta}^{\text{I}}_{\text{e}}}
\end{array}\right]\notag\\&=\frac{2T}{\sigma^2_0}\left[\begin{array}{c|c}
\mathbf{J}_{11} & \mathbf{J}_{12}\\
\hline \mathbf{J}^T_{12}&\mathbf{J}_{22}\\
\end{array}\right]
\end{align}
where $J_{xy}=\text{Re}\left\{\frac{\partial \mathbf{u}^H}{\partial \bm{\xi}_i}\frac{\partial \mathbf{u}}{\partial \bm{\xi}_j}\right\}$ for $x,y\in\left\{\theta_{\text{e}},r_{\text{e}}\right\}$.  With (\ref{partial_derivative}) at hand, we have
\begin{subequations}\label{FIM}
	\begin{align}
&J_{xy}=|\tilde\beta_{\text{e}}|^2\text{Re}\left(\dot{\mathbf{G}}_{y}\mathbf{R}\dot{\mathbf{G}}^H_{x}\right),\\
&J_{x\tilde{\beta}^{\text{R}}_{\text{e}}}=\tilde{\beta}^*_{\text{e}} \text{Re}\left(\text{Tr}(\tilde{\mathbf{ G}}\mathbf{R}\dot{\mathbf{G}}^H_{x})\right),\\
&J_{x\tilde{\beta}^{\text{I}}_{\text{e}}}=j\tilde{\beta}^*_{\text{e}} \text{Re}\left(\text{Tr}(\tilde{\mathbf{ G}}\mathbf{R}\dot{\mathbf{G}}^H_{x})\right),\\
&J_{\tilde{\beta}^{\text{R}}_{\text{e}}\tilde{\beta}^{\text{R}}_{\text{e}}}=J_{\tilde{\beta}^{\text{I}}_{\text{e}}\tilde{\beta}^{\text{I}}_{\text{e}}}=\text{Tr}\left(\tilde{\mathbf{ G}}\mathbf{R}\tilde{\mathbf{ G}}^H\right).
	\end{align}
\end{subequations}
By substituting (\ref{FIM}) into (\ref{FIM_2}), we can obtain $\mathbf{J}_{11}$, $\mathbf{J}_{12}$, and $\mathbf{J}_{22}$.

\section{Reformulation of the CRB Matrix}
According to (\ref{FIM_3}) and $\mathbf{R}=\mathbf{PP}^H$, we can derive
\begin{align}
\mathbf{J}^{-1}_{22}=\frac{1}{\text{Tr}\left(\tilde{\mathbf{ G}}\mathbf{PP}^H\tilde{\mathbf{ G}}^H\right)}\mathbf{I}_2
\label{inverse_J11}
\end{align}
and 
\begin{align}\label{J_12}
\mathbf{J}_{12}\mathbf{J}^T_{12}=&\left[\begin{array}{cc}
\text{Re}\left(\tilde{\beta}^*_{\text{e}}\kappa_1\right) & -\text{Im}\left(\tilde{\beta}^*_{\text{e}}\kappa_1\right)\\
\text{Re}\left(\tilde{\beta}^*_{\text{e}}\kappa_2\right)&-\text{Im}\left(\tilde{\beta}^*_{\text{e}}\kappa_2\right)\\
\end{array}\right]\times\notag\\&\left[\begin{array}{cc}
\text{Re}\left(\tilde{\beta}^*_{\text{e}}\kappa_1\right) & 
\text{Re}\left(\tilde{\beta}^*_{\text{e}}\kappa_2\right)\\-\text{Im}\left(\tilde{\beta}^*_{\text{e}}\kappa_1\right)&-\text{Im}\left(\tilde{\beta}^*_{\text{e}}\kappa_2\right)\\
\end{array}\right]\notag\\
=&\left|\tilde{\beta}_{\text{e}}\right|^2\left[\begin{array}{cc}
\left|\kappa_1\right|^2 &\text{Re}\left(\kappa_1\kappa^*_2\right)\\
\text{Re}\left(\kappa_1\kappa^*_2\right)&\left|\kappa_2\right|^2\\
\end{array}\right]
\end{align}
where the symbol $\times$ denotes matrix multiplication between two matrices, $\kappa_1=\text{Tr}\left(\tilde{\mathbf{ G}}\mathbf{PP}^H\dot{\mathbf{G}}^H_{\theta_{\text{e}}}\right)$ and $\kappa_2=\text{Tr}\left(\tilde{\mathbf{ G}}\mathbf{PP}^H\dot{\mathbf{G}}^H_{r_{\text{e}}}\right)$, respectively.

Based on (\ref{inverse_J11}), we obtain
\begin{align}
\mathbf{J}_{12}\mathbf{J}^{-1}_{22}\mathbf{J}^T_{12}=\frac{1}{\text{Tr}\left(\tilde{\mathbf{ G}}\mathbf{PP}^H\tilde{\mathbf{ G}}^H\right)}\mathbf{J}_{12}\mathbf{J}^T_{12}
\label{J12}
\end{align}
Substituting (\ref{J12}) and (\ref{J_12}) into (\ref{CRB}), we can obtain

\begin{align}
\text{CRB}\left(r_{\text{e}},\theta_{\text{e}}\right)
= 
\frac{\sigma^2_{\text{e}}\text{Tr}\!\left(\tilde{\mathbf{ G}}\mathbf{PP}^H\tilde{\mathbf{ G}}^H\right)}
{2T\left|\tilde{\beta}_{\text{e}}\right|^2}
\left[\begin{array}{c|c}
C_{11} & C_{12}\\
\hline C_{21}&C_{22}\\
\end{array}\right]^\dagger,
\end{align}
where
\begin{subequations}\label{FIM22}
	\begin{align}
C_{11}=&\text{Tr}\!\left(\dot{\mathbf{G}}_{\theta_{\text{e}}}\mathbf{PP}^H\dot{\mathbf{G}}^H_{\theta_{\text{e}}}\right)
 \text{Tr}\!\left(\tilde{\mathbf{ G}}\mathbf{PP}^H\tilde{\mathbf{ G}}^H\right)  \notag\\
 &-\left|\text{Tr}\!\left(\tilde{\mathbf{ G}}\mathbf{PP}^H
\dot{\mathbf{G}}^H_{\theta_{\text{e}}}\right)\right|^2,\\
C_{12}=&\text{Re}\!\left(\text{Tr}\!\left(\dot{\mathbf{G}}_{r_{\text{e}}}\mathbf{PP}^H
 \dot{\mathbf{G}}^H_{\theta_{\text{e}}}\right)
 \text{Tr}\!\left(\tilde{\mathbf{ G}}\mathbf{PP}^H\tilde{\mathbf{ G}}^H\right) \right.\notag \\
 &\left.-\text{Tr}\!\left(\tilde{\mathbf{ G}}\mathbf{PP}^H
 \dot{\mathbf{G}}^H_{\theta_{\text{e}}}\right)
 \text{Tr}\!\left(\tilde{\mathbf{ G}}\mathbf{PP}^H
\dot{\mathbf{G}}^H_{r_{\text{e}}}\right)^*\right),\\
C_{21}=&\text{Re}\!\left(\text{Tr}\!\left(\dot{\mathbf{G}}_{\theta_{\text{e}}}\mathbf{PP}^H
 \dot{\mathbf{G}}^H_{r_{\text{e}}}\right)
 \text{Tr}\!\left(\tilde{\mathbf{ G}}\mathbf{PP}^H\tilde{\mathbf{ G}}^H\right)\right. \notag\\
 &\left.-\text{Tr}\!\left(\tilde{\mathbf{ G}}\mathbf{PP}^H
 \dot{\mathbf{G}}^H_{\theta_{\text{e}}}\right)
 \text{Tr}\!\left(\tilde{\mathbf{ G}}\mathbf{PP}^H
\dot{\mathbf{G}}^H_{r_{\text{e}}}\right)^*\right),\\
C_{22}=&\text{Tr}\!\left(\dot{\mathbf{G}}_{r_{\text{e}}}\mathbf{PP}^H\dot{\mathbf{G}}^H_{r_{\text{e}}}\right)
 \text{Tr}\!\left(\tilde{\mathbf{ G}}\mathbf{PP}^H\tilde{\mathbf{ G}}^H\right)\notag \\
 &-\left|\text{Tr}\!\left(\tilde{\mathbf{ G}}\mathbf{PP}^H
\dot{\mathbf{G}}^H_{r_{\text{e}}}\right)\right|^2.
	\end{align}
\end{subequations}
When the target distance $r_{\text{e}}$ is known, $\dot{\mathbf{G}}_{r_{\text{e}}}=0$, and thus $\text{CRB}\left(r_{\text{e}},\theta_{\text{e}}\right)$ reduces to a matrix with only the upper-left element being nonzero, from which (\ref{Error_angle}) holds directly. Conversely, when the target angle is known, $\dot{\mathbf{G}}_{\theta_{\text{e}}}=0$, and $\text{CRB}\left(r_{\text{e}},\theta_{\text{e}}\right)$ reduces to a matrix with only the lower-right element being nonzero, yielding (\ref{Error_distance}).

\section{Proof of Lemma 1}
Since $\mathbf{X}$ is Hermitian positive semi-definite, we have
\begin{align}
\text{Re}\left(\text{Tr}(\mathbf{A}\mathbf{X})\right)
= \text{Tr}\left(\frac{\mathbf{A}+\mathbf{A}^H}{2}\mathbf{X}\right)
= \text{Tr}(\mathbf{S}\mathbf{X}).
\end{align}
By the eigen decomposition $\mathbf{S}=\mathbf{U}\mathbf{D}\mathbf{U}^H$ with real diagonal $\mathbf{D}$, decompose
\begin{align}
\mathbf{D} = \mathbf{D}_+ - \mathbf{D}_-,
\end{align}
where $\mathbf{D}_+\succeq\mathbf{0}$ collects the nonnegative eigenvalues and $\mathbf{D}_-\succeq\mathbf{0}$ collects the absolute values of the negative ones. Hence,
\begin{align}
\mathbf{S}=\mathbf{U}\mathbf{D}_+\mathbf{U}^H - \mathbf{U}\mathbf{D}_-\mathbf{U}^H.
\end{align}
Substituting into $\text{Tr}(\mathbf{S}\mathbf{X})$ yields (\ref{DC}). Thus, Lemma 1 can be proved.

	\ifCLASSOPTIONcaptionsoff
	\newpage
	\fi
	
	\bibliographystyle{IEEEtran}
	\bibliography{references}
    
\end{document}